\documentclass[12pt]{iopart}
\usepackage{iopams}
\newtheorem{defi}{Definition}
\newtheorem{lem}[defi]{Lemma}
\newtheorem{thm}[defi]{Theorem}
\newtheorem{cor}[defi]{Corollary}
\newtheorem{prop}[defi]{Proposition}
\newtheorem{ex}{Example}
\newtheorem{rem}{Remark}
\def\QED{\mbox{\rule[0pt]{1.5ex}{1.5ex}}}

\def\endproof{\hspace*{\fill}~\QED\par\endtrivlist\unskip}

\def\Tr{\mathop{\rm Tr}\nolimits}

\def\SU{\mathop{\rm SU}\nolimits}
\def\SL{\mathop{\rm SL}\nolimits}

\def\Id{\mathop{\rm I}\nolimits}

\def\real{\mathbb{R}}
\def\complex{\mathbb{C}}

\def\Label#1{\label{#1}\ [\ #1\ ]\ }
\def\Label{\label}
\begin{document}
\jl{1}
\title[Two quantum analogues of Fisher information]
{Two quantum analogues of Fisher information from \\
a large deviation viewpoint of quantum estimation}
\author{Masahito Hayashi\dag}
\address{\dag\
Laboratory for Mathematical Neuroscience, 
Brain Science Institute, RIKEN\\
2-1, Hirosawa, Wako, Saitama 351-0198, Japan\\
e-mail masahito@brain.riken.go.jp}
\begin{abstract}
We discuss two quantum analogues of Fisher 
information, symmetric logarithmic derivative (SLD)
Fisher information and 
Kubo-Mori-Bogoljubov (KMB) Fisher information
from a large deviation viewpoint of
quantum estimation and prove that
the former gives the true bound
and the latter gives the bound of 
consistent superefficient estimators.
In another comparison,
it is shown that the difference between them
is characterized by the 
change of the order of limits.
\end{abstract}
\submitted
\maketitle
\section{Introduction}\Label{intro}
Fisher information ont only plays a central role in 
statisical inference,
but also
coincides with a natural inner product 
in a distribution family.
It is defined as
\begin{eqnarray}
J_{\theta}:= \int_{\Omega} l_{\theta}(\omega)^2 p_{\theta}(\omega)\,d \omega,
\quad
l_{\theta}(\omega)p_\theta(\omega)  =
\frac{\,d p_{\theta}(\omega)}{\,d \theta}
\Label{I1}
\end{eqnarray}
for a 
probability distribution family $\{p_\theta | \theta \in \Theta
\subset \real\}$ with a probability space $\Omega$.
However, the quantum version of Fisher information cannot be uniquely
determined.
In general, there is a serious arbitrariness
concerning the order among non-commutative observables
in the quantization of products of several variables.
The problem of the arbitrarity of the quantum version of Fisher
information is due to the same reason.
The geometrical properties of its quantum analogues 
have been discussed 
by many authors\cite{Na}\cite{P10}\cite{Petz}\cite{Petz2}.

One quantum analogue is the Kubo-Mori-Bogoljubov (KMB)
Fisher information $\tilde{J}_{\rho}$ defined by
\begin{eqnarray}
\tilde{J}_{\theta}:= \int_0^1 \Tr \rho_{\theta}^{t} \tilde{L_{\theta}}
\rho_{\theta}^{1-t}\tilde{L_{\theta}} \,d t,
\quad
\int_0^1  \rho_{\theta}^{t} \tilde{L_{\theta}} \rho_{\theta}^{1-t}
\,d t = \frac{\,d \rho_{\theta}}{\,d \theta} \Label{f1}
\end{eqnarray}
for a quantum state family $\{\rho_\theta
\in {\cal S}({\cal H})| \theta \in \Theta \}$,
where ${\cal S}({\cal H})$ is the set of density matrixes on ${\cal H}$
and the Hilbert space ${\cal H}$ corresponds to
the physical system of
interest\cite{Na}\cite{P10}\cite{Petz}\cite{Petz2}.
As proven in \ref{as0}, it can be characterized as 
the limit of quantum relative entropy, which 
plays an important role in several topics of
quantum information theory, for example,
quantum channel coding \cite{HoCh}\cite{SW}, 
quantum source 
coding \cite{Schumacher}\cite{Winter}\cite{Hm} 
and quantum hypothesis testing \cite{HP}\cite{ON}.
Moreover, as mentioned in section \ref{ss2}, 
this inner product is closely related to the canonical correlation
of the linear response theory in statistical mechanics\cite{KTH}.
As mentioned in \ref{rem1},
it appears to be the most natural quantum extension 
from an information geometrical viewpoint.
Thus, one might expect that it is 
significant in quantum estimation,
but its estimation-theoretical characterization 
has not been sufficiently clarified.

Another quantum analogue is symmetric logarithmic 
derivative (SLD) 
Fisher information
\begin{eqnarray}
J_{\theta} := \Tr L_{\theta}^2 \rho_{\theta}, \quad
\frac{1}{2}(L_{\theta}\rho_{\theta}+ \rho_{\theta}L_{\theta}) 
=\frac{\,d \rho_{\theta}}{\,d \theta},
\Label{a9} 
\end{eqnarray}
where $L_\theta$ is called the symmetric logarithmic 
derivative\cite{Hel67}.
It is closely related to the achievable lower bound of 
mean square error (MSE)
not only for the one-parameter case
\cite{Hel67}\cite{Hel}\cite{HolP}, 
but also for the multi-parameter case
\cite{GillM}\cite{2}\cite{3}
in quantum estimation.
The difference between the two can be regarded as
the difference in the order of the operators,
and reflects the two ways of defining 
Fisher information for a probability distribution family.

Currently, the former is closely related to the 
quantum information theory
while the latter is related to the quantum estimation theory.
These two inner products 
have been discussed only in separate contexts. 
In this paper, to clarify the difference between them,
we introduce a large deviation viewpoint of 
quantum estimation as a unified viewpoint,
whose classical version was initiated by
Bahadur\cite{Ba60}\cite{Ba67}\cite{Bahadur}.
This method may not be conventional in 
mathematical statistics,
but seems a suitable setting for 
a comparison between two quantum analogues from 
an estimation viewpoint.
This type of comparison was initiated 
by Nagaoka \cite{Nag}\cite{Na1},
and is discussed in further depth in this paper.
Such a large deviation evaluation of 
quantum estimation is closely related to
the exponent of the overflow probability 
of quantum universal variable-length 
coding\cite{HM2002}.

This paper is structured as follows:
Before we state the main results,
we review the classical estimation theory
including Bahadur's large deviation theory,
which has been done in section \ref{ss1}.
After this review,
we briefly outline the main results in section \ref{ss2},
i.e., the difference is characterized from
three contexts.
To simplify the notations, even if we need the Gauss notation $[~]$,
we omit it when this does not cause confusion.
Some proofs are very complicated and are presented
in the appendixes.

\section{Review of classical estimation theory}\Label{ss1}
We review the relationship
 between the parameter estimation for the
probability distribution family $\{p_\theta | \theta \in \Theta
\subset \real\}$ with a probability space $\Omega$
and its Fisher information.
The definition of Fisher information is given 
not only by (\ref{I1}), but also by the 
limit of the 
relative entropy (Kullback-Leibler divergence)
$D(p \|q ):=
\int_{\Omega} \left(\log p(\omega) - \log
q(\omega) \right)p(\omega)\,d \omega$ as
\begin{eqnarray}
J_{\theta}:=  \lim_{\epsilon \to 0} \frac{2}{\epsilon^2}
D(p_{\theta+\epsilon}\|p_{\theta})
\Label{I2}.
\end{eqnarray}
These two definitions (\ref{I1}) and (\ref{I2})
coincide under some regularity conditions for a family.

Next, we consider a map $f$ from $\Omega$ to $\Omega'$.
Similarly to other information quantities,
(for example Kullback divergence etc)
the inequality
\begin{eqnarray}
J_{\theta} \ge J_{\theta} '
\Label{j7}
\end{eqnarray}
holds, where
$J_{\theta} '$ is Fisher information 
of the family 
$\{p_\theta \circ f^{-1}| \theta \in \Theta\}$.
Inequality (\ref{j7}) is called the monotonicity.
According to \v{C}encov\cite{Cencov},
any information quantities satisfying (\ref{j7})
coincide with a constant times
Fisher information $J_\theta$.

For an estimator that is defined as a map from the data set $\Omega$ to
the parameter set $\Theta$, we sometimes consider the {\it unbiasedness}
condition:
\begin{eqnarray}
\int_{\Omega} T(\omega) p_{\theta}(\omega)\,d \omega
= \theta , \quad \forall \theta \in \Theta.
\Label{I3}
\end{eqnarray}
The MSE of any
unbiased estimator $T$ is evaluated by 
the following inequality 
(Cram\'{e}r-Rao inequality),
\begin{eqnarray}
\int_{\Omega} (T(\omega)-\theta)^2 p_{\theta}(\omega)\,d \omega
\ge \frac{1}{J_{\theta}},
\Label{I4}
\end{eqnarray}
which follows from 
Schwartz inequality with respect to (w.r.t.) 
the inner product 
$\langle X, Y \rangle:=\int_{\Omega} 
X(\omega) Y(\omega) p_{\theta}(\omega)\,d \omega$ 
for variables $X,Y$.
When the number of data $
\vec{\omega}_n := (\omega_1 , \ldots,\omega_n)$, which obeys the
unknown probability $p_\theta$, is sufficiently 
large, we discuss a
sequence $\{ T_n\}$ of estimators $T_n(\vec{\omega}_n)$.
If $\{ T_n\}$ is suitable as a sequence of estimators,
we can expect that it converges to the true parameter $\theta$ in
probability, i.e., it satisfies the {\it weak consistency} condition:
\begin{eqnarray}
\lim_{n \to \infty} p_{\theta}^n
\{ | T_n  - \theta | \,> \epsilon \} = 0 , \quad
\forall \epsilon \,> 0,  \forall \theta \in \Theta.
\Label{I5}
\end{eqnarray}

Usually, the performance of a sequence $\{ T_n\}$ of estimators
is measured by the speed of its convergence.
As one criterion, we focus on the speed of the convergence in 
MSE.
If a sequence $\{ T_n\}$ of estimators satisfies the 
weak consistency condition and some regularity 
conditions, the asymptotic version of 
Cram\'{e}r-Rao inequality,
\begin{eqnarray}
\lim_{n \to \infty} n 
\int_{\Omega} (T_n(\vec{\omega}_n)-\theta)^2 p_{\theta}^n(\omega)\,d \omega
\ge \frac{1}{J_{\theta}}
\Label{I6},
\end{eqnarray}
holds.
If it satisfies only the weak consistency condition, 
it is possible that it surpasses
the bound of (\ref{I6}) at a specific subset.
Such a sequence of estimators is called superefficient.
We can reduce its error to any amount at a specific subset 
with the measure $0$
under the weak consistency condition (\ref{I5}).

As another criterion, we evaluate the decreasing rate of the tail
probability:
\begin{eqnarray}
\beta(\{ T_n \}, \theta, \epsilon ):=
\lim_{n \to \infty} \frac{-1}{n}
\log p_{\theta}^n \{ | T_n - \theta| \,> \epsilon \}.
\Label{I7}
\end{eqnarray}
This method was initiated by 
Bahadur\cite{Ba60}\cite{Ba67}\cite{Bahadur},
and was a much discussed topic among mathematical
statisticians in the 1970's. From 
the monotonicity of the divergence, we can prove the inequality 
\begin{eqnarray}
\beta(\{ T_n \}, \theta, \epsilon )
\le \min \{ D( p_{\theta+ \epsilon} \| p_{\theta}) ,
D( p_{\theta- \epsilon} \| p_{\theta})\}
\Label{I8}
\end{eqnarray}
for any weakly consistent sequence $\{ T_n \}$ of estimators.
Its proof is essentially given in our proof of Theorem \ref{T1}.
Since it is difficult to analyze $\beta(\{ T_n \},\theta,\epsilon)$
except in the case of an exponential family,
we focus on another quantity $\alpha( \{ T_n \}, \theta):=
\lim_{\epsilon \to 0} \frac{1}{\epsilon^2}\beta(\{ T_n \}, \theta,
\epsilon )$. For an exponential family,
see \ref{s10}. Taking the limit $\epsilon \to +0$,
we obtain the inequality
\begin{eqnarray}
\alpha( \{ T_n \}, \theta) \le \frac{J_{\theta}}{2}
\Label{I9}.
\end{eqnarray}
If $T_n$ is the maximum likelihood estimator (MLE),
the equality of (\ref{I9}) holds under some regularity conditions
for the family \cite{Bahadur} \cite{Fu}.
This type of discussion is different from the MSE type 
of discussion 
in deriving (\ref{I9}) from only
the weak consistency condition. Therefore, there is no
consistent 
superefficient estimator w.r.t.\ the large deviation
evaluation.

Indeed, we can relate the above large deviation type of 
discussion in the estimation to Stein's lemma 
in simple hypothesis testing as follows.
In simple hypothesis testing, we decide whether
the null hypothesis should be accepted or rejected from the 
data $\vec{\omega}_n := (\omega_1 , \ldots,
\omega_n)$ which obeys an unknown probability. For the decision, 
we must define an 
{\it acceptance region} $A_n$ as a subset of $\Omega^n$.
If the null hypothesis is $p$ and the alternative is $q$, the first
error (though the true distribution is $p$, we reject the null hypothesis)
probability $\beta_{1,n}(A_n)$
and the second error (though the true distribution is $q$, we accept the null
hypothesis)
probability $\beta_{2,n}(A_n)$ are given by
\begin{eqnarray*}
\beta_{1,n}(A_n):= 1- p^n(A_n), \quad
\beta_{2,n}(A_n):= q^n(A_n).
\end{eqnarray*}
Regarding the decreasing rate of the second error probability under the 
constant constraint of the first error probability,
the equation 
\begin{eqnarray}
\lim_{n \to \infty} \frac{-1}{n}
\log \min \{ \beta_{2,n}(A_n) | \beta_{1,n}(A_n)\le \epsilon\}
= D( p \| q )
, \quad \epsilon \,> 0 \Label{I10}
\end{eqnarray}
holds (Stein's lemma).
Inequality (\ref{I8}) can be derived from this lemma. We can regard
the large deviation type of evaluation in the estimation to 
be Stein's lemma in the case where the null hypothesis is close to
the alternative one.

\section{Outline of main results}\Label{ss2}
Let us return to the quantum case.
In a quantum setting, 
we focus two quantum analogues
of Fisher information,
KMB Fisher information $\tilde{J}_{\theta}$ and 
SLD Fisher information $J_{\theta}$.
Indeed, if the state $\rho_{\theta}$ is nondegenerate,
SLD $L_\theta$ is not uniquely determined.
However, as is proven in \ref{as1}, 
SLD Fisher information $J_\theta$ is uniquely determined, 
i.e., it is 
independent of the choice of the SLD $L_\theta$.

On the other hand,
according to Chap. 7 in Amari and Nagaoka \cite{Na},
$\tilde{L_{\theta}}$ has another form 
\begin{eqnarray}
\tilde{L_{\theta}}= \frac{\,d \log \rho_{\theta}}{\,d \theta}.
\Label{j2}
\end{eqnarray}
As is proven by using formula (\ref{j2}) in \ref{as0}, 
KMB Fisher information $\tilde{J}_{\theta}$
can be characterized as 
the limit of the the quantum relative entropy 
$D(\rho\|\sigma):= \Tr \rho( \log \rho - \log \sigma)$
in the following way
\begin{eqnarray}
  \tilde{J}_{\theta}=
\lim_{\epsilon\to 0}\frac{2}{\epsilon^2}
D(\rho_{\theta+\epsilon} \|\rho_{\theta})
\Label{I12}.
\end{eqnarray}
Moreover, in the linear response theory of statistical physics,
given an equilibrium state $\rho$,
when a variable $A$ fluctuates with a small value $\delta$,
another variable $B$ also is thought to fluctuate with a constant 
times $\delta$ \cite{KTH}.
Its coefficient is called the canonical correlation 
and given by
\begin{eqnarray}
\int_0^1 \Tr \rho_{\theta}^{t} (A-\Tr \rho A)
\rho_{\theta}^{1-t}(B-\Tr \rho B) \,d t.
\end{eqnarray}
Thus, KMB Fisher information $\tilde{J}_{\theta}$
is thought to be more natural from a viewpoint 
of statistical physics.

As another quantum analogue, 
right logarithmic derivative
(RLD) Fisher information $\check{J_{\theta}}$:
\begin{eqnarray*}
\check{J}_{\theta}:= \Tr \rho_{\theta} \check{L}_{\theta}
\check{L}_{\theta}^*, \quad
\rho_\theta\check{L_{\theta}}
=\frac{\,d \rho_{\theta}}{\,d \theta}
\end{eqnarray*}
is known.
When $\rho_\theta$ does not commute 
$\frac{\,d \rho_{\theta}}{\,d \theta}$ and $\rho_\theta \,> 0$,
the RLD $\check{L_\theta}$ is not self-adjoint.
Since it is not useful in the one-parameter case,
we do not discuss it in this paper.
Since the difference in definitions can be regarded as 
the difference in the order of operators,
these quantum analogues
coincide when all states of the family are 
commutative with each other.
However, in the general case, they do not coincide
and the inequality $\tilde{J}_{\theta} \ge J_{\theta}$
holds, as exemplified in section \ref{s1}.
Concerning some information-geometrical properties,
see \ref{rem1}.

In the following, we consider how the roles these quantum analogues of 
Fisher information play in the parameter estimation for the state 
family.
As is discussed in detail in section \ref{s1},
the estimator is described by the pair of 
positive operator valued measure (POVM) $M$ 
(which corresponds to the measurement and is defined in section \ref{s1})
and the map from the data set to the parameter space $\Theta$.
Similarly to the classical case, we can define an unbiased estimator.
For any unbiased estimator $E$, the SLD Cram\'{e}r-Rao inequality 
\begin{eqnarray}
V(E) \ge \frac{1}{J_{\theta}}
\Label{I13}
\end{eqnarray}
holds, where $V(E)$ is the mean square error (MSE) of the estimator
$E$.

In an asymptotic setting,
as a quantum analogue of the $n$-i.i.d.\ condition, 
we treat the quantum $n$-i.i.d.\ condition, i.e.,
we consider the case where
the number of systems independently 
prepared in the same unknown state is sufficiently 
large, in section \ref{s2}.
In this case, the measurement is denoted by a POVM $M^n$ on the
composite system ${\cal H}^{\otimes n}$ and the state is described by the
tensor product density matrix $\rho^{\otimes n}$. Of course, such POVMs 
include a POVM that requires quantum correlations 
between the respective quantum systems
in the measurement apparatus.
Similarly to the classical case, for a sequence $\vec{E}=\{ E^n\}$
of estimators, we can define the {\it weak consistency} 
condition given in (\ref{a1}).
In mathematical statistics, 
the square root $n$ consistency,
local asymptotic minimax theorems
and Bayesian theorem are important topics as the
asymptotic theory, but
it seems too difficult to link these quantum settings 
and KMB Fisher information $\tilde{J}_{\theta}$.
Thus, in this paper, in order to compare two quantum analogues 
from a unified framework,
we adopt Bahadur's large deviation theory as follows.
As is discussed in section
\ref{s2}, we can similarly define the quantities
$\beta(\vec{E},\theta,\epsilon),\alpha(\vec{E},\theta)$.
Similarly to (\ref{I8})(\ref{I9}),
under the weak consistency (WC) condition, the inequalities
\begin{eqnarray}
\beta(\vec{E},\theta,\epsilon) &\le&
\min\{ D( \rho_{\theta+\epsilon}\|\rho_{\theta}),
D( \rho_{\theta-\epsilon}\|\rho_{\theta})\} 
\nonumber \\
\alpha(\vec{E},\theta) &\le& \frac{1}{2}\tilde{J}_{\theta}
\Label{I14}
\end{eqnarray}
hold. From these discussions, the bound in the large deviation
type of evaluation seems different from the one in the 
MSE case.
However, as mentioned in section \ref{s3}, the inequality
\begin{eqnarray}
\alpha(\vec{E},\theta) \le \frac{1}{2}J_{\theta}
\Label{I15}
\end{eqnarray}
holds if the sequence $\vec{E}$ satisfies the strong
consistency
(SC) condition introduced in section \ref{s3} as 
a stronger condition.
As is mentioned in section \ref{s5},
these bounds can be attained in their respective senses.
Therefore, roughly speaking,
the difference between the two quantum analogues
can be regarded as the difference 
in consistency conditions and can be characterized as
\begin{eqnarray*}
\sup_{\vec{E}:\hbox{SC}}
\lim_{\epsilon \to 0}
\frac{1}{\epsilon^2}\beta(\vec{E},\theta,\epsilon)
&= \frac{1}{2}J_{\theta} \\
\sup_{\vec{E}:\hbox{WC}}
\lim_{\epsilon \to 0}
\frac{1}{\epsilon^2}\beta(\vec{E},\theta,\epsilon)
&= \frac{1}{2}\tilde{J}_{\theta} .
\end{eqnarray*}
Even if we restrict our estimators to strongly consistent ones,
the difference between two appears as
\begin{eqnarray}
\sup_{\vec{M}:\hbox{SC}}
\liminf_{\epsilon \to 0}
\frac{1}{\epsilon^2}
\beta(\vec{M},\theta,\epsilon)
&= \frac{J_{\theta}}{2} \Label{Dec1}\\
\liminf_{\epsilon \to 0}
\frac{1}{\epsilon^2}
\sup_{\vec{M}:\hbox{SC}}
\beta(\vec{M},\theta,\epsilon)
&= \frac{\tilde{J}_{\theta}}{2}\Label{Dec2},
\end{eqnarray}
where, for a precise statement, as expressed in section \ref{s9},
we need more complicated definitions.

However, we should consider that the bound 
$\frac{J_{\theta}}{2}$ is more meaningful for the following two reasons.
The first reason is the fact that we can construct the sequence of
estimators attaining the bound $\frac{J_{\theta}}{2}$ at all
points, which is proven in section \ref{s5}.
On the other hand, there is a sequence of estimators attaining the bound 
$\frac{\tilde{J_{\theta}}}{2}$ at one point $\theta$, but 
it cannot attain the bound at all points.
The other reason is the naturalness of the conditions 
for deriving the bound $\frac{J_{\theta}}{2}$.
In other words, an estimator attaining $\frac{J_{\theta}}{2}$
is natural, but an estimator attaining $\frac{\tilde{J_{\theta}}}{2}$
is very irregular.
Such a sequence of estimators can be regarded as a 
consistent superefficient
estimator and does not satisfy regularity conditions other 
than the weak consistency condition.
This type of discussion of the superefficiency
is different from the MSE type of discussion in that any
consistent superefficient estimator is bounded by 
inequality (\ref{I14}). 

To consider the difference between the two quantum analogues
of Fisher information in more details,
we must analyze how we can achieve the bound 
$\frac{\tilde{J}_\theta}{2}$.
It is important in this analysis to 
consider the relationship between the above discussion 
and the quantum version of Stein's lemma
in simple hypothesis testing.
Similarly to the classical case, when the null hypothesis is the
state $\rho$ and the alternative is the state $\sigma$, we evaluate
the decreasing rate of the second error probability under the constant 
constraint $\epsilon \,> 0$ of the first error probability. 
As was proven in quantum Stein's lemma,
its exponential 
component is given by the quantum relative entropy 
$D(\rho\|\sigma)$ for any $\epsilon\,> 0$. 
Hiai and Petz \cite{HP} constructed a sequence of tests 
to attain the optimal rate
$D(\rho\|\sigma)$,
by constructing the sequence $\{M^n\}$ of POVMs
such that 
\begin{eqnarray}
\lim_{n \to \infty}
\frac{1}{n}D({\rm P}^{M^n}_{\rho}\|{\rm P}^{M^n}_{\sigma} )=D(\rho\|\sigma).
\Label{I16}
\end{eqnarray}
Ogawa and Nagaoka \cite{ON} proved that there is 
no test exceeding the
bound $D(\rho\|\sigma)$.
It was proven by Hayashi \cite{Haya} that by
using the group representation theory,
we can construct the POVM satisfying (\ref{I16}) 
independently of $\rho$.
For the reader's convenience, we give
a review of this in \ref{s6}.
As discussed in section \ref{s52},
this type of construction is useful for the construction 
of an estimator attaining
the bound $\frac{\tilde{J}_{\theta}}{2}$
at one point.
Since the proper bound of the large deviation is
$\frac{J_{\theta}}{2}$, we cannot regard the quantum estimation 
as the limit of the quantum Stein's lemma.

In order to consider the properties of estimators
attaining the bound $\frac{\tilde{J}_{\theta}}{2}$ at one point
from another viewpoint,
we consider the restriction that makes
such a construction impossible.
We introduce a class of estimators whose 
POVMs do not require a quantum correlation 
in the quantum apparatus in section \ref{s30}.
In this class, we assume that the 
POVM on the $l$-th system 
is chosen from $l-1$ data. 
We call such an estimator an adaptive estimator.
When an adaptive estimator $\vec{E}$ satisfies 
the weak consistency condition, the inequality
\begin{eqnarray}
\alpha(\vec{E},\theta) \le \frac{1}{2}J_{\theta}
\Label{I17}
\end{eqnarray}
holds (See section \ref{s3}). 
Similarly, we can define a class of estimators that use
quantum correlations up to $m$ systems.
We call such an estimator an $m$-adaptive estimator.
For any $m$-adaptive weakly consistent
estimator $\vec{E}$, inequality
(\ref{I17}) holds.
Therefore, it is impossible to construct a sequence of estimators 
attaining the bound $\frac{\tilde{J}_{\theta}}{2}$ if we
fix the number of systems in which we use quantum 
correlations. As mentioned in section \ref{s30}, 
taking limit $m \to \infty$, we obtain
\begin{eqnarray}
\lim_{m \to \infty}
\lim_{\epsilon \to 0}
\sup_{\vec{M}:m\hbox{-AWC}}\frac{1}{ \epsilon^2}
\beta(\vec{M},\theta,\epsilon)=
\frac{J_\theta}{2}\Label{H52a} ,
\end{eqnarray}
where $m$-AWC denotes an $m$-adaptive weakly consistent
estimator.
However, as the third characterization of the difference
between the two quantum analogues,
as precisely mentioned in section \ref{s9},
the equation
\begin{eqnarray}
\lim_{\epsilon \to 0}
\lim_{m \to \infty}
\sup_{\vec{M}:m\hbox{-ASC}}\frac{1}{ \epsilon^2}
\beta(\vec{M},\theta,\epsilon)=
\frac{\tilde{J}_\theta}{2} \Label{H51a}
\end{eqnarray}
holds, where $m$-ASC denotes an $m$-adaptive strongly consistent
estimator.
A more narrow class of estimators
is treated in equation (\ref{H51a}) 
than in equation (\ref{Dec2}).
Equations (\ref{H52a}) and (\ref{H51a}) indicate that
the order of limits $\lim_{m \to\infty}$ and $\lim_{\epsilon \to 0}$
is more crucial than the difference between two types
of consistencies.

\begin{rem}\Label{rem2}\rm
In the estimation only of the spectrum of 
a density matrix in a unitary-invariant family,
the natural inner product in the parameter space
is unique and equals Fisher inner product
in the distribution family
whose element is the probability distribution corresponding to 
eigenvalues of a density matrix.
In addition, the achievable bound is derived by Keyl and Werner
 \cite{KW},
and coincides with the bound uniquely given 
by the above inner product.
For detail, see \ref{ap15}.
\end{rem}
\section{Review of non-asymptotic setting in quantum estimation}\Label{s1}
In a quantum system, in order to discuss the 
probability distribution which the data obeys,
we must define a POVM.

A POVM $M$ is defined as a map from Borel sets of the data set $\Omega$ 
to the set of bounded, self-adjoint and positive semi-definite operators,
which satisfies
\begin{eqnarray*}
M(\emptyset )= 0, \quad M(\Omega)= \Id,\quad
\sum_i M(B_i) = M ( \cup B_i) \hbox{ for disjoint sets}.
\end{eqnarray*}
If the state on the quantum system ${\cal H}$ is 
a density operator $\rho$
and we perform a measurement
corresponding to a POVM $M$ on the system,
the data obeys the probability distribution ${\rm P}^M_{\rho}(B):=
\Tr \rho M(B)$.
If a POVM $M$ satisfies $M(B)^2= M(B)$ for any Borel set $B$,
$M$ is called a projection-valued measure (PVM).
The spectral measure of a self-adjoint operator $X$
is a PVM, and is denoted by $E(X)$.
For $1 \,> \lambda \,> 0$ and any POVMs $M_1$ and $M_2$ 
taking values in $\Omega$, the POVM $B \mapsto \lambda 
M_1(B) + (1-\lambda)M_2(B)$ is called the {\it random 
combination} of $M_1 $ and $M_2$ in the 
ratio $\lambda : 1- \lambda$.
Even if $M_1$'s data set $\Omega_1$ is different from 
$M_2$'s data set $\Omega_2$,
$M_1$ and $M_2$ can be regarded as POVMs taking values in
the disjoint union set $\Omega_1 \coprod \Omega_2:=
(\Omega_1 \times \{1\}) \cup (\Omega_2 \times \{2\})$.
In this case, we can define a random combination of 
$M_1$ and $M_2$ as a POVM taking values in $\Omega_1 
\coprod \Omega_2$ and call it the {\it disjoint random 
combination}. In this paper, we simplify the probability 
${\rm P}^M_{\rho_{\theta}}$ and the relative entropies 
$D(\rho_{\theta_0} \|\rho_{\theta_1})$ and  
$D({\rm P}^{M}_{\rho_{\theta_0}} 
\|{\rm P}^{M}_{\rho_{\theta_1}})$ to 
${\rm P}^M_{\theta}$, $D(\theta_0\|\theta_1)$ and 
$D^M(\theta_0\|\theta_1)$, respectively.

In the one-parameter quantum estimation, the estimator
is described by a pair comprising a POVM 
and a map from its data set to the real number set $\real $.
Since the POVM $M \circ T^{-1}$ takes values in the real number set
$\real $, we can regard any estimator as a POVM taking values in 
the real number set $\real $.
In order to evaluate MSE,
Helstrom \cite{Hel67,Hel} derived the
SLD Cram\'{e}r-Rao inequality as a quantum counterpart of 
Cram\'{e}r-Rao inequality (\ref{G14}).
If an estimator $M$ satisfies 
\begin{eqnarray}
\int_\real x \Tr \rho_{\theta} M(\,d x) = \theta , \quad
\forall \theta \in \Theta , \Label{G31}
\end{eqnarray}
it is called unbiased.
If $\theta-\theta_0$ is sufficiently small,
we can obtain the following approximation in the 
neighborhood of $\theta_0$:
\begin{eqnarray*}
\int_\real x \Tr \rho_{\theta_0}  M( \,d x) 
+ \left(\int_\real x 
\Tr \left. \frac{\partial \rho_{\theta}}{\partial \theta}
\right|_{\theta= \theta_0} M( \,d x) \right)(\theta - \theta_0 )
\cong \theta_0 +  (\theta - \theta_0 ).
\end{eqnarray*}
It implies the following two conditions:
\begin{eqnarray}
\int_\real x 
\Tr \left. \frac{\partial \rho_{\theta}}{\partial \theta}
\right|_{\theta= \theta_0} M( \,d x) &=&1 \Label{G32}\\
\int_\real x \Tr \rho_{\theta_0}  M( \,d x) &=& \theta_0. \Label{G33}
\end{eqnarray}
If an estimator $M$ satisfies (\ref{G32}) and (\ref{G33}),
it is called locally unbiased at $\theta_0$.
For any locally unbiased estimator $M$ (at $\theta$),
the inequality, which is called the SLD Cram\'{e}r-Rao inequality,
\begin{eqnarray}
\int_\real (x- \theta) ^2 \Tr \rho_{\theta} M( \,d x) \ge 
\frac{1}{J_{\theta}}
\Label{G14}
\end{eqnarray}
holds.
Similarly to the classical case,
this inequality is derived from the Schwartz inequality
with respect to SLD Fisher information
$\langle X | Y \rangle:= 
\Tr \rho_{\theta}\frac{XY + YX}{2}$
\cite{Hel67} \cite{Hel} \cite{HolP}.

The equality of (\ref{G14}) holds when
the estimator is given by the spectral decomposition 
$E(\frac{L_\theta}{J_{\theta}}+\theta)$
of $\frac{L_\theta}{J_{\theta}}+\theta$, where $L_\theta$ is the SLD 
at $\theta$ and is defined by (\ref{a9}).
This implies that
SLD Fisher information $J_{\theta_0}$ 
coincides with 
Fisher information at $\theta_0$ of the probability family
$\left \{\left.
 {\rm P}^{E(\frac{L_{\theta_0}}{J_{\theta_0}}+\theta_0)}_{\theta}
\right| \theta \in \Theta\right\}$.
The monotonicity of quantum relative entropy \cite{6.1}
\cite{6}
gives the following evaluation 
of  the probability family
$\left \{\left.
 {\rm P}^{E(\frac{L_{\theta_0}}{J_{\theta_0}}+\theta_0)}_{\theta}
\right| \theta \in \Theta\right\}$:
\begin{eqnarray*}
D^{E\left(\frac{L_{\theta_0}}{J_{\theta_0}}+\theta_0\right)}
(\theta\|\theta_0)
\le
D( \theta \| \theta_0).
\end{eqnarray*}
Taking the limit $\theta \to \theta_0$,
we have
\begin{eqnarray}
J_\theta \le \tilde{J}_\theta \Label{G12}.
\end{eqnarray}
In this paper,
we discuss inequality (\ref{G12}) from the viewpoint
of the large deviation type of evaluation of the quantum 
estimation. 
The following families are treated 
as simple examples of the one-parameter 
quantum state family, in the latter.
\begin{ex}\rm
\noindent{\bf [One-parameter 
equatorial 
spin 1/2 system state family]}:
\begin{eqnarray*}
{\cal S}_r := 
\left \{\left .
\rho_{\theta}:= 
\frac{1}{2}\left (
\begin{array}{cc}
1 + r \cos \theta & r \sin \theta \\
r \sin \theta & 1- r \cos \theta
\end{array}
\right)
\right| 
0 \le \theta \,< 2 \pi \right\}
\end{eqnarray*}
In this family, we calculate 
\begin{eqnarray*}
D( \rho_{\theta} \| \rho_0) =
\frac{r}{2}( 1- \cos \theta)\log \frac{1+ r}{1-r} \\
\tilde{J}_{\theta}=
\frac{r}{2}\log \frac{1+ r}{1-r}\\
J_{\theta}= r^2 .
\end{eqnarray*}
Since the relations 
$\tilde{J}_{\theta}=\infty$ and $J_{\theta}= 1$ hold 
in the case of $r=1$, 
the two quantum analogues are completely different.
\end{ex}
\begin{ex}\rm
\noindent{\bf [One-parameter quantum Gaussian state family
and half-line quantum Gaussian state family]}:
We define the boson coherent vector $| \alpha \rangle:=
e^{- \frac{|\alpha|^2}{2}}
\sum_{n=0}^{\infty}\frac{\alpha^{n}}{\sqrt{n !}}|n \rangle$,
where $|n \rangle$ is the number vector on $L^2(\real)$.
The quantum Gaussian state is defined as
\begin{eqnarray*}
\rho_{\theta}:=
\frac{1}{\pi \overline{N}} \int_{\complex}
| \alpha \rangle \langle \alpha |
e^{-\frac{|\alpha-\theta|^2}{\overline{N}}}\,d^2 \alpha
, \quad \forall \theta \in \complex.
\end{eqnarray*}
We call $\{ \rho_{\theta}| \theta \in \real\}$
the {\it one-parameter quantum Gaussian state family},
and call $\{ \rho_{\theta}| \theta \ge 0 ( \theta \in \real^+=
[0, \infty))\}$ the
{\it half-line quantum Gaussian state family}.
In this family, we can calculate
\begin{eqnarray}
D( \rho_{\theta} \| \rho_{\theta_0} ) =
\log \left( 1+ \frac{1}{\overline{N}}\right)|\theta-\theta_0|^2, \nonumber \\
\tilde{J}_{\theta} =
2 \log \left( 1+ \frac{1}{\overline{N}}\right), \nonumber \\
J_{\theta} =
\frac{2}{\overline{N} +\frac{1}{2}}.\nonumber
\end{eqnarray}
\end{ex}
\section{The bound under the weak consistency condition}\Label{s2}
We introduce the quantum 
independent-identical density (i.i.d.)
condition in order to treat an asymptotic setting.
Suppose that $n$-independent physical systems are
prepared in the same state $\rho$.
Then, the quantum state of the composite system is described 
by
\begin{eqnarray*}
\rho^{\otimes n} :=  \underbrace{\rho \otimes \cdots \otimes \rho }_{n}
\hbox{ on } {\cal H}^{\otimes n} ,
\end{eqnarray*}
where the tensor product space ${\cal H}^{\otimes n}$ is defined by
\begin{eqnarray*}
{\cal H}^{\otimes n} := 
\underbrace{{\cal H} \otimes \cdots \otimes {\cal H}}_{n} .
\end{eqnarray*}
We call this condition the quantum i.i.d.\ condition,
which is a quantum analogue of the independent-identical
distribution condition.
In this setting,
any estimator is described by a
POVM $M^n$ on ${\cal H}^{\otimes n}$, whose data set is $\real$.
In this paper,
we simplify ${\rm P}^{M^n}_{\rho_{\theta}^{\otimes n}}$ 
and $D({\rm P}^{M^n}_{\rho_{\theta_0}^{\otimes n}} \|
{\rm P}^{M^n}_{\rho_{\theta_1}^{\otimes n}})$
to ${\rm P}^{M^n}_{\theta}$ and $ D^{M^n}(\theta_0\|\theta_1)$.
The notation $M \times n$ denotes 
the POVM in which we perform 
the POVM $M$ for the respective $n$ systems.

\begin{defi}\rm 
{\bf [Weak consistency condition]}:
A sequence of estimators 
$\vec{M}:= \{ M^n \}_{n=1}^\infty$ is called 
{\it weakly consistent} if 
\begin{eqnarray}
\lim_{n\to \infty}
{\rm P}^{M^n}_{\theta}
\left\{ | \hat{\theta}- \theta| \,> \epsilon\right\} =0, \quad
\forall \theta \in \Theta, \forall \epsilon \,> 0, \Label{a1}
\end{eqnarray}
where $\hat{\theta}$ is the estimated value.
\end{defi}
This definition means that 
the estimated value $\hat{\theta}$ 
converges to the true value $\theta$ in probability,
and can be regarded as the quantum 
extension of (\ref{I5}).

Now, we focus on the exponential component of the tail probability
as follows:
\begin{eqnarray*}
\beta(\vec{M},\theta, \epsilon):= 
\limsup_{n\to \infty} \frac{-1}{n}\log {\rm P}^{M^n}_{\theta}
\left\{ | \hat{\theta}- \theta| \,> \epsilon\right\}.
\end{eqnarray*}
We usually discuss 
the following value instead of $\beta(\vec{M},\theta, \epsilon)$
\begin{eqnarray}
\alpha(\vec{M},\theta):= 
\limsup_{\epsilon \to 0} \frac{1}{\epsilon^2}
\beta(\vec{M},\theta, \epsilon) \Label{a8}
\end{eqnarray}
because it is too difficult to discuss $\beta(\vec{M},\theta,
\epsilon) $.
The following theorem can be proven from the monotonicity
of the quantum relative entropy.
\begin{thm}[Nagaoka\cite{Nag,Na1}]\Label{T1}
If a POVM $M^n$ on ${\cal H}^{\otimes n}$ satisfies the 
weakly consistent condition (\ref{a1}),
the inequalities
\begin{eqnarray}
\beta(\vec{M},\theta, \epsilon) &\le& 
\inf \{ D(\rho_{\theta'}\| \rho_{\theta}) |
|\theta - \theta' | \,< \epsilon \} \Label{a21} \\
\alpha(\vec{M},\theta) &\le& \frac{\tilde{J_{\theta}}}{2}\Label{a3}
\end{eqnarray}
hold.
\end{thm}
Even if the parameter set $\Theta$ is not open 
(e.g., the closed half-line $\real^+ := [0 , \infty )$),
this theorem holds.

\begin{proof}
The monotonicity of the quantum relative entropy
yields the inequality
\begin{eqnarray*}
D(\rho^{\otimes n}_{\theta'}\| \rho^{\otimes n}_{\theta})
\ge 
p_{n,\theta'}\log \frac{p_{n,\theta'}}{p_{n,\theta}}
+
\left(1-p_{n,\theta'}\right)
\log \frac{1-p_{n,\theta'}}{1-p_{n,\theta}},
\end{eqnarray*}
for any $\theta'$ satisfying $
| \theta' - \theta| \,> \epsilon$,
where we denote 
the probability ${\rm P}^{M^n}_{\theta''}
\left\{ | \hat{\theta}- \theta| \,> \epsilon\right\}$
by $p_{n,\theta''}$.
Using the inequality $-\left(1-p_{n,\theta'}\right)
\log \left(1-p_{n,\theta}\right) \ge 0$,
we have
\begin{eqnarray}
- \frac{\log {\rm P}^{M^n}_{\theta}
\left\{ | \hat{\theta}- \theta| \,> \epsilon\right\} }
{n} 
= - \frac{\log p_{n,\theta}}{n} \le 
\frac{D(\rho^{\otimes n}_{\theta'}\| \rho^{\otimes n}_{\theta})
+ h(p_{n,\theta'})}{n p_{n,\theta'}} ,\Label{D12}
\end{eqnarray}
where $h$ is the binary entropy defined by 
$h(x):= - x \log x - (1-x)\log (1-x)$.
Since the assumption guarantees that $p_{n,\theta'} \to 1$,
the inequality
\begin{eqnarray}
\beta(\vec{M},\theta, \epsilon) \le D(\rho_{\theta'}\| \rho_{\theta})
\Label{a2}
\end{eqnarray}
holds, where we use the additivity of quantum relative entropy:
\begin{eqnarray*}
D(\rho^{\otimes n}_{\theta'}\| \rho^{\otimes n}_{\theta})
= n D(\rho_{\theta'}\| \rho_{\theta}).
\end{eqnarray*}
Thus, we obtain (\ref{a21}).
Taking the limit $\epsilon \to 0$ in inequality 
(\ref{a2}), we obtain (\ref{a3}).
\end{proof}
As another proof, we can prove this inequality 
as a corollary of the quantum 
Stein's lemma \cite{HP,ON}.
\section{The bound under the strong consistency condition}\Label{s3}
As discussed in section \ref{s1}, 
the SLD Cram\'{e}r-Rao inequality guarantees
that the lower bound of MSE is given by SLD Fisher information.
Therefore, it is expected that 
the bound is connected with SLD Fisher information
for large deviation.
In order to discuss the relationship between SLD Fisher information and 
the bound for large deviation,
we need another characterization with respect to 
the limit of the tail probability.
We thus define 
\begin{eqnarray}
\underline{\beta}(\vec{M},\theta, \epsilon)&:= &
\liminf_{n\to \infty} \frac{-1}{n}\log {\rm P}^{M^n}_{\theta}
\left\{ | \hat{\theta}- \theta| \,> \epsilon\right\} \nonumber\\
\underline{\alpha}(\vec{M},\theta)&:= &
\liminf_{\epsilon \to 0} \frac{1}{\epsilon^2}
\underline{\beta}(\vec{M},\theta, \epsilon)  \Label{a22}.
\end{eqnarray}

In the following, we attempt to link the quantity 
$\underline{\alpha}(\vec{M},\theta)$ with SLD Fisher 
information.
For this purpose, it is suitable to focus on an information 
quantity
that satisfies the additivity and the monotonicity,
as in the proof of Theorem 1.
Its limit should be SLD Fisher information.
The Bures distance $b(\rho,\sigma):=
\sqrt{2(1- \Tr | \sqrt{\rho}\sqrt{\sigma}|)}=
\sqrt{\min_{U:unitary} \Tr ( \sqrt{\rho}- \sqrt{\sigma}U)
( \sqrt{\rho}- \sqrt{\sigma}U)^*}$
is known to be an information quantity
whose limit is SLD Fisher information, as mentioned in Lemma \ref{L2}.
Of course, it can be regarded as a quantum analogue of 
the Hellinger distance,
and satisfies the monotonicity.

\begin{lem}[Uhlmann \cite{Uhlmann}, Matsumoto \cite{Matu}]\Label{L2}
If there exists an SLD $L_{\theta}$ satisfying (\ref{a9}),
then the equation 
\begin{eqnarray}
\frac{1}{4}J_{\theta}=
\lim_{\epsilon \to 0}
\frac{b^2( \rho_{\theta} ,\rho_{\theta+\epsilon})}{\epsilon^2}
\Label{a18}
\end{eqnarray}
holds.
\end{lem}
A proof of Lemma \ref{L2} is given in \ref{as1}.
As discussed in the latter, 
the Bures distance satisfies the monotonicity.
Unfortunately, 
the Bures distance does not satisfy the additivity.

However, the quantum affinity
$I(\rho \| \sigma):= -8 \log \Tr \left| \sqrt{\rho}\sqrt{\sigma}\right|=
-8 \log \left(1- \frac{1}{2}b(\rho,\sigma)^2 \right)$
satisfies the additivity:
\begin{eqnarray}
I( \rho^{\otimes n} \| \sigma^{\otimes n})=
n I( \rho \| \sigma).\Label{a4}
\end{eqnarray}
Its classical version is called affinity 
in the following form\cite{LC}:
\begin{eqnarray}
I( p\| q)= -8 \log \left(\sum_{i} \sqrt{p_i}\sqrt{q_i}\right). \Label{a7}
\end{eqnarray}
As a trivial deformation of (\ref{a18}),
the equation 
\begin{eqnarray}
\lim_{\epsilon \to 0}\frac{ I(\rho_{\theta}\|\rho_{\theta+\epsilon} )}
{\epsilon^2}= J_{\theta} \Label{a7.1}
\end{eqnarray}
holds.
The quantum affinity satisfies the monotonicity
w.r.t.\ any measurement $M$ 
(Jozsa \cite{Jozsa}, Fuchs \cite{Fuchs}):
\begin{eqnarray}
I( \rho \| \sigma)  \ge I\left(\left. {\rm P}^{M}_\rho 
\right\| {\rm P}^{M}_\sigma \right)
= - 8 \log \sum_{\omega} \left( \sqrt{{\rm P}^{M}_\rho(\omega)}
\sqrt{{\rm P}^{M}_\sigma( \omega)}\right). \Label{A7}
\end{eqnarray}
The most simple proof of (\ref{A7}) is given by Fuchs 
\cite{Fuchs} who directly proved that
\begin{eqnarray}
\Tr \sqrt{\sqrt{\rho}\sigma\sqrt{\rho}}
\le \sum_{\omega} \left(\sqrt{{\rm P}^{M}_\rho(\omega)}
\sqrt{{\rm P}^{M}_\sigma( \omega)}\right).\Label{A7.1}
\end{eqnarray}
For the reader's convenience,
a proof of (\ref{A7.1}) is given in \ref{apb}. From 
(\ref{a4}),(\ref{a7.1}) and (\ref{A7}), we can expect 
that SLD Fisher information is, in a sense, closely 
related to a large deviation type of bound. From 
the additivity and the monotonicity of the quantum affinity,
we can show the following lemma.
\begin{lem}\Label{L1}
The inequality 
\begin{eqnarray}
4
\inf_{\{ s | 1 \ge s \ge 0\}}
\left( \underline{\beta}' (\vec{M},\theta, s \delta )+
\underline{\beta}' (\vec{M},\theta+ \delta , (1-s)\delta)
\right) \le
I( \rho_{\theta} \| \rho_{\theta+\delta}) \Label{a19}
\end{eqnarray}
holds, where 
we define $\underline{\beta}' (\vec{M},\theta, \delta ):=
\lim_{\epsilon \to +0}
\underline{\beta} (\vec{M},\theta, \delta- \epsilon)$.
\end{lem}
A proof of Lemma \ref{L1} is given in \ref{as2}.
However, Lemma \ref{L1} 
cannot yield an inequality w.r.t.\ $\alpha(\vec{M},\theta)$
under the weak consistency condition,
unlike inequality (\ref{a2}).
Therefore, we consider a stronger condition,
which is given in the following.
\begin{defi}\Label{defsc}\rm
\noindent{\bf [Strong consistency condition]}:
A sequence of estimators $\vec{M}= \{M^n \}_{n=1}^{\infty}$
is called {\it strongly consistent}
if the convergence of (\ref{a22}) is uniform for the parameter
$\theta$ and if
$\underline{\alpha}( \vec{M}, \theta)$ is continuous for $\theta$.
A sequence of estimators is called {\it strongly consistent at $\theta$}
if there exists a neighborhood $U$ of $\theta$ such that
it is strongly consistent in $U$.
\end{defi}
The square root $n$ consistency is familiar 
in the field of  mathematical statistics.
However, in the large deviation setting,
this strong consistency seems more suitable
than the square root $n$ consistency.

As a corollary of Lemma \ref{L1}, we have the following theorem.
\begin{thm}\Label{T2}
Assume that there exists the SLD $L_{\theta}$ satisfying (\ref{a9}).
If a sequence of estimators $\vec{M}= \{M^n \}_{n=1}^{\infty}$
is strongly consistent at $\theta$, then the inequality 
\begin{eqnarray}
\underline{\alpha}( \vec{M}, \theta) \le \frac{J_{\theta}}{2} 
\Label{a11}
\end{eqnarray}
holds.
\end{thm}

\begin{proof} From the above assumption,
for any real $\epsilon \,> 0$ and any element 
$\theta \in \Theta$, there exists a sufficiently 
small real $\delta \,> 0$
such that $(\underline{\alpha}(\vec{M},\theta)- \epsilon) {\epsilon'}^2
\le \underline{\beta}'(\vec{M},\theta,\epsilon'),
\underline{\beta}'(\vec{M},\theta+\delta,\epsilon')$
for $\forall \epsilon' \,< \delta$.
Therefore, inequality (\ref{a19}) yields 
the relations
\begin{eqnarray}
&&2 (\underline{\alpha}(\vec{M},\theta)- \epsilon) \delta^2 =
4 ( \underline{\alpha}( \vec{M}, \theta) - \epsilon) 
\inf_{\{ s| 1 \ge s \ge 0\}}
\left( s^2 \delta^2 + (1-s)^2 \delta^2\right) \nonumber\\
&& \le 
4 \inf_{\{ s| 1 \ge s \ge 0\}}
\left( \underline{\beta}' (\vec{M},\theta, s \delta )+
\underline{\beta}' (\vec{M},\theta+ \delta , (1-s)\delta)\right)
\le
I(\rho_{\theta} \| \rho_{\theta+\delta}). \Label{a10}
\end{eqnarray}
Lemma \ref{L2} and (\ref{a10})
guarantee 
(\ref{a11}) for $\forall \theta \in \Theta$.
\end{proof}
\begin{rem}\rm
Inequality (\ref{A7.1}) can be regarded as a special case of the 
monotonicity w.r.t.\
any trace-preserving CP (completely positive) map $C: {\cal S}({\cal H}_1 )\to 
{\cal S}({\cal H}_2)$: 
\begin{eqnarray}
\left(\Tr \left| \sqrt{\rho}\sqrt{\sigma}\right| \right)^2\le 
\left(\Tr \left| \sqrt{C(\rho)}\sqrt{C(\sigma)}\right|\right)^2 \Label{h12}
\end{eqnarray}
which is proven by Jozsa \cite{Jozsa}
because the map $\rho \mapsto {\rm P}^M_{\rho}$ can be regarded as a 
trace-preserving CP map 
from the $C^*$ algebra of bounded operators on ${\cal H}$
to the commutative $C^*$ algebra $C(\Omega)$,
where $\Omega$ is the data set.
\end{rem}

\section{Attainabilities of the bounds}\Label{s5}
Next, we discuss the attainabilies of the two bounds
$\tilde{J}_{\theta}$ and $J_{\theta}$ in their 
respective senses.
In this section, we discuss the attainabilies 
in two cases:
the first case is the one-parameter quantum Gaussian state family,
and the second case is an arbitrary one-parameter finite-dimensional quantum state family
that satisfies some assumptions.
\subsection{One-parameter quantum Gaussian state family}\Label{s51}
In this subsection, we discuss the attainabilies
in the one-parameter quantum Gaussian state family.

\begin{thm}\Label{T2.1}
In the one-parameter quantum Gaussian state family,
the sequence of estimators $\vec{M^s}=
\{ M^{s,n} \}_{n=1}^{\infty}$ (defined in the following) satisfies 
the strong consistency condition and 
the relations
\begin{eqnarray}
\alpha ( \vec{M^s}, \theta)
=\underline{\alpha} ( \vec{M^s}, \theta)
= \frac{J_{\theta}}{2}=
\frac{1}{\overline{N} +\frac{1}{2}}
. \Label{D5}
\end{eqnarray}
\end{thm}
\noindent{\bf [Construction of $\vec{M^s}$]}:
We perform the POVM $E(Q)$ for all systems,
where $Q$ is the position operator 
on $L^2(\real)$.
The estimated value $\xi_n$ 
is determined to be the mean value of $n$ data.
\endproof

\begin{proof}
Since the equation
\begin{eqnarray*}
{\rm P}_{| \alpha \rangle \langle \alpha|}^{E(Q)}(\,d x)=
\sqrt{\frac{2}{\pi}}e^{- 2(x- \alpha_x)^2} \,d x
\end{eqnarray*}
holds, we have the equation
\begin{eqnarray*}
{\rm P}_{\theta}^{E(Q)}(\,d x)=
\left( {\rm P}_{\rho_{\theta}}^{E(Q)}(\,d x)
\right)=
\frac{1}{\pi N}
\int_{\complex} 
{\rm P}_{| \alpha \rangle \langle \alpha|}^{E(Q)}(\,d x)
e^{-\frac{|\alpha-\theta|^2}{N}}\,d^2 \alpha \\
=
\sqrt{\frac{2}{\pi(2\overline{N}+1)}}
e^{-\frac{2(x- \theta)^2}{2\overline{N}+1}}\,d x.
\end{eqnarray*}
Thus, we obtain the equation 
\begin{eqnarray*}
{\rm P}_{\theta}^{M^{s,n}}(\,d \xi_n)=
\sqrt{\frac{2}{\pi(2\overline{N}+1)n}}
e^{-\frac{2(\xi_n- \theta)^2}{(2\overline{N}+1)n}}\,d 
\xi_n,
\end{eqnarray*}
which implies that
\begin{eqnarray}
\beta(\vec{M^s},\theta,\epsilon)=
\lim \frac{-1}{n}\log
{\rm P}^{M^{s,n}}_\theta
\{ | \xi_n - \theta| \,> \epsilon \}
= \frac{\epsilon^2}{\overline{N} +\frac{1}{2}}.
\Label{4-14}
\end{eqnarray}
Therefore, the sequence of estimators
$\vec{M^s}= \{ M^{s,n} \}_{n=1}^{\infty}$
attains the bound $\frac{J_{\theta}}{2}$
and satisfies the strong consistency condition.
\end{proof}
\begin{prop}\Label{P1}
In the half-line quantum Gaussian state family,
the sequence of estimators $\vec{M^w}= \{ M^{w,n}\}_{n=0}^{\infty}$
(defined in the following)
satisfies the weak consistency condition and
the strong consistency condition at $\real^+ \setminus \{ 0 \}$
and the relations 
\begin{eqnarray}
\underline{\alpha}(\vec{M^w},0)&= &
\alpha(\vec{M^w},0)=\frac{\tilde{J}_{0}}{2}=
\log \left( 1+ \frac{1}{\overline{N}}\right),
\Label{D37}\\
\underline{\alpha}(\vec{M^w},\theta) &= &
\alpha(\vec{M^w},\theta)
= \frac{J_{\theta}}{2}=
\frac{1}{\overline{N} +\frac{1}{2}}, \quad \forall \theta \in \real^+ \setminus
\{ 0 \}.\Label{D39}
\end{eqnarray}
\end{prop}
This proposition indicates
the significance of the uniformity of the convergence of
(\ref{a22}).
This proposition 
is proven in \ref{asc}.

\noindent{\bf [Construction of $\vec{M^w}$]}:
We perform the following unitary evolution:
\begin{equation*}
\rho_{\theta}^{\otimes n} \mapsto
\rho_{\sqrt{n}\theta}\otimes \rho_0^{\otimes (n-1)} .
\end{equation*}
For detail, see \ref{s11}.
We perform the number measurement $E(N)$ of the first system
whose state is $\rho_{\sqrt{n}\theta}$, and let $k$ be its 
data,
where the number operator $N$ is defined as $N:=
\sum_n n | n \rangle \langle n|$.
The estimated value $T_n$ is determined by $T_n:= \sqrt{\frac{k}{n}}$.
\endproof

\begin{thm}\Label{4-26}
In the one-parameter quantum Gaussian state family,
for any $\theta \in \real$,
the sequence of estimators $\vec{M^w_{\theta_1}}=
\{ M^{w,n}_{\theta_1} \}_{n=1}^{\infty}$ (defined in the following)
satisfies 
the weak consistency condition and 
the relations
\begin{eqnarray}
\underline{\alpha}(\vec{M^w_{\theta_1}},\theta_1)= 
\alpha(\vec{M^w_{\theta_1}},\theta_1)
= \frac{\tilde{J}_{\theta}}{2}=
\log \left( 1+ \frac{1}{\overline{N}}\right)
\Label{D6}.
\end{eqnarray}
\end{thm}
\noindent{\bf [Construction of $\vec{M^w_{\theta_1}}$]}:
We divide $n$ systems into two groups.
One consists of $\sqrt{n}$ systems and the other, of 
$n- \sqrt{n}$ systems.
We perform the PVM $E(Q)$ for every system
in the first group. 
Let $\xi_{\sqrt{n}}$ be 
the mean value in the first group,
i.e., we perform the PVM $M^{s,\sqrt{n}}$
for the first system.
At the second step, we perform the following unitary evolution for the second group.
\begin{equation*}
\rho_{\theta}^{\otimes (n-\sqrt{n})} \mapsto
\rho_{\theta-\theta_1}^{\otimes (n-\sqrt{n})}
\end{equation*}
For details, see \ref{s11}.
We perform the POVM $M^{w,n-\sqrt{n}}$ for 
the system whose state is
$\rho_{\theta-\theta_1}^{\otimes (n-\sqrt{n})}$;
the data is written as $T_{n-\sqrt{n}}$.
Then, we decide the final estimated value $\hat{\theta}$
as
\begin{equation*}
\hat{\theta}:=
\theta_1 + {\rm sgn} (\xi_{\sqrt{n}}- \theta_1) T_{n- \sqrt{n}}.
\end{equation*}
\endproof

\begin{proof}
Since
\begin{eqnarray*}
{\rm P}_{\theta_1}^{M^{w,n}_{\theta_1}}
\left\{ \left|\hat{\theta}- \theta_1
\right|\,> \epsilon\right\}
=
{\rm P}_{0}^{M^{w,n-\sqrt{n}}}
\left\{\left|T_{n-\sqrt{n}}\right|\,> \epsilon
\right\},
\end{eqnarray*}
we have
\begin{eqnarray*}
\beta(\vec{M_{\theta_1}^w},\theta_1)=
\lim \frac{-1}{n} \log 
{\rm P}_{\theta_1}^{M^{w,n}_{\theta_1}}
\left\{ \left|\hat{\theta}- \theta_1
\right|\,> \epsilon\right\}\\
= 
\lim \frac{n-\sqrt{n}}{n} 
\frac{-1}{n- \sqrt{n}}\log 
{\rm P}_{0}^{M^{w,n-\sqrt{n}}}
\left\{\left|T_{n-\sqrt{n}}\right|\,> \epsilon
\right\}
= \beta(\vec{M^w},0).
\end{eqnarray*}
As is shown in \ref{asc}, we have
\begin{equation*}
\beta(\vec{M^w},0)= \epsilon^2 
\log \left( 1+ \frac{1}{\overline{N}}\right),
\end{equation*}
which implies (\ref{D6}).
Next, we prove the consistency in the case 
where $\theta \,> \theta_1$.
In this case, it is sufficient to discuss
the case where $\theta - \theta_1 \,> \epsilon \,>0$.
Since the first measurement 
$M^{s,\sqrt{n}}$ and the second 
one $M^{w,n-\sqrt{n}}$ are performed 
independently, we obtain
\begin{eqnarray*}
\fl {\rm P}_{\theta}^{M^{w,n}_{\theta_1}}
\left\{ \left|\hat{\theta}- \theta_1
\right|\,> \epsilon\right\}
\le
{\rm P}_{\theta}^{M^{w,n-\sqrt{n}}}
\left\{ \left| 
T_{n-\sqrt{n}} - (\theta - \theta_1)
\right|\,> \epsilon\right\} 
+
{\rm P}_{\theta}^{M^{s,\sqrt{n}}}
\left\{ \xi_{\sqrt{n}} - \theta_1
\le 0 \right\}.
\end{eqnarray*}
Proposition \ref{P1} guarantees
that the first term goes to $0$,
and Theorem \ref{T2.1} guarantees
that the second term goes to $0$.
Thus, we obtain the consistency of 
$\vec{M^w_{\theta_1}}$.
Similarly, we can prove the weak consistency the case where
$\theta \,< \theta_1$.
\end{proof}
\subsection{Finite dimensional family}\Label{s52}
In this subsection, 
we treat the case where the dimension of the Hilbert space ${\cal H}$ is $k$ (finite).
As for the attainability of the RHS of inequality
(\ref{a11}), we have the following lemma.

\begin{lem}\Label{L3}
Let $\theta_0$ be fixed in $\Theta$.
Under Assumptions 1 and 2,
the sequence of estimators
$\vec{M}_{\theta_0}^s$ (defined in the following) satisfies
the strong consistency condition at $\theta_0$ 
(defined in Def. \ref{defsc}) and the
relation
\begin{eqnarray}
\alpha ( \vec{M^s_{\theta_0}}, \theta_0)
=\underline{\alpha} ( \vec{M^s_{\theta_0}}, \theta_0)
= \frac{J_{\theta_0}}{2}. \Label{H54}
\end{eqnarray}
\end{lem}
\par\noindent
{\bf [Assumption 1]}:
The map $\theta \mapsto \rho_{\theta}$ is $C^1$ and 
$\rho_{\theta} \,> 0$.
\par\noindent
{\bf [Assumption 2]}:
The map $\theta \mapsto \Tr \rho_{\theta}
\frac{L_{\theta_0}}{J_{\theta_0}}$ is injective
i.e., one-to-one.
\par\noindent
{\bf [Construction of $\vec{M^s_{\theta_0}}$]}:
We perform the POVM $E(\frac{L_{\theta_0}}{J_{\theta_0}})$
for all systems. 
The estimated value is determined to be the mean value plus
$\theta_0$.
\endproof

\noindent{\it Proof of Lemma \ref{L3}:}\quad From Assumption 2,
the weak consistency is satisfied.
Let $\delta \,> 0$ be a sufficiently small number.
Define 
the function 
\begin{eqnarray}
\phi_{\theta,\theta_0}(s):=
\Tr \rho_{\theta} \exp \left( s 
\left( \frac{L_{\theta_0}}{J_{\theta_0}} -
 \frac{\Tr \rho_{\theta} L_{\theta_0}}{J_{\theta_0}}
\right) \right). \Label{14-3}
\end{eqnarray}
Since $\left\| \frac{L_{\theta_0}}{J_{\theta_0}}\right\| \,< \infty$
and $\Tr \rho_\theta \left( \frac{L_{\theta_0}}{J_{\theta_0}} -
 \frac{\Tr \rho_{\theta} L_{\theta_0}}{J_{\theta_0}}
\right)= 0$,
we have
\begin{eqnarray*}
\lim_{s \to 0} \frac{\phi_{\theta,\theta_0}(s)-1}{s^2}= 
\frac{1}{2} \Tr \rho_{\theta} 
\left( \frac{L_{\theta_0}}{J_{\theta_0}} -
 \frac{\Tr \rho_{\theta} L_{\theta_0}}{J_{\theta_0}}
\right)^2 .
\end{eqnarray*}
When $\| \theta- \theta_0 \| $ is sufficiently small,
the function
$x \to 
\sup_s (x  s - \log \phi_{\theta,\theta_0}(s))$
is continuous in $(-\delta ,\delta)$.
Using Cram\'{e}r's theorem \cite{Buck},
we have
\begin{eqnarray*}
\fl \lim_{n \to \infty}
\frac{-1}{n} \log 
{\rm P}_{\theta}^{M^{s,n}_{\theta_0}}
\left\{ | \hat{\theta}- \theta_0| \,> \epsilon \right\}
=
\min \left\{ \sup_s (\epsilon s - 
\log \phi_{\theta,\theta_0}(s)),
 \sup_{s'}(-\epsilon s' 
- \log \phi_{\theta,\theta_0}(s'))\right\} 
\end{eqnarray*}
for $\epsilon \,< \delta$.
Taking the limit $\epsilon \to 0$,
we have
\begin{eqnarray*}
&\lim_{\epsilon \to 0}
\lim_{n \to \infty}\frac{-1}{\epsilon^2 n}
{\rm P}_{\theta_0}^{M^{s,n}_{\theta_0}}
\{ | \hat{\theta}- \theta_0| \,> \epsilon \}\\
&=
\min \left\{ 
\lim_{\epsilon \to 0}
\frac{\sup_s (\epsilon s - \log \phi_{\theta,\theta_0}(s))}
{\epsilon^2},
\lim_{\epsilon \to 0}
\frac{\sup_s (-\epsilon s - \log \phi_{\theta,\theta_0}(s))}
{\epsilon^2} \right\}
=
\frac{1}{2}c_{\theta,\theta_0}^{-1},
\end{eqnarray*}
where
\begin{eqnarray*}
c_{\theta,\theta_0} := 
\Tr \rho_{\theta} 
\left( \frac{L_{\theta_0}}{J_{\theta_0}} -
 \frac{\Tr \rho_{\theta} L_{\theta_0}}{J_{\theta_0}}
\right)^2 
\end{eqnarray*}
because 
\begin{eqnarray*}
\fl \epsilon s- \log \phi_{\theta,\theta_0}(s)
\cong \epsilon s - \log ( 1+ \frac{1}{2}c_{\theta,\theta_0} s^2)
\cong \epsilon s - \frac{1}{2}c_{\theta,\theta_0} s^2
= - \frac{c_{\theta,\theta_0}}{2}\left( s- \frac{\epsilon}{c_{\theta,\theta_0}}\right)
+ \frac{\epsilon^2}{2 c_{\theta,\theta_0}}.
\end{eqnarray*}
The above convergence is uniform for the neighborhood 
of $\theta_0$.
Taking the limit $\theta \to \theta_0$,
we have
\begin{eqnarray*}
\lim_{\theta \to \theta_0}
\Tr \rho_{\theta} 
\left( \frac{L_{\theta_0}}{J_{\theta_0}} -
 \frac{\Tr \rho_{\theta} L_{\theta_0}}{J_{\theta_0}}
\right) ^2
=
J_{\theta_0}^{-1}
=
\Tr \rho_{\theta_0} 
\left( \frac{L_{\theta_0}}{J_{\theta_0}} -
 \frac{\Tr \rho_{\theta_0} L_{\theta_0}}{J_{\theta_0}}
\right)^2.
\end{eqnarray*}
Thus, we can check (\ref{H54}) and the strong consistency 
in the neighborhood of $\theta_0$.
\endproof
However, this sequence of estimators $\vec{M^s_{\delta}}$
depends on the true parameter $\theta_0$.
We should construct a sequence of estimators
that satisfies the strong consistency condition
and attains the bound $\frac{J_{\theta_0}}{2}$ at all 
points $\theta_0$.
Since such a construction is too difficult, we introduce another strong consistency condition
that is weaker than the above and under which 
inequality (\ref{a11}) holds.
We construct a sequence of estimators
that satisfies this strong consistency condition and 
attains the bound given in (\ref{a11}) for all $\theta$ in a weak sense.

\noindent {\bf [Second strong consistency condition]}:
A sequence of estimators $\vec{M}= \{ M^n\}$ is called 
second strongly consistent if there exists a sequence of functions
$\{ \underline{\beta}_m ( \vec{M}, \theta, \epsilon)\}_{m=1}^\infty$
such that
\begin{itemize}
\item $\displaystyle
\lim_{m \to \infty}\lim_{\epsilon \to 0}\frac{1}{\epsilon^2}
\underline{\beta}_m(\vec{M},\theta,\epsilon)=
\underline{\alpha}(\vec{M},\theta)$.
\item $\displaystyle\lim_{\epsilon \to 0}\frac{1}{\epsilon^2}
\underline{\beta}_m(\vec{M},\theta,\epsilon)\le
\underline{\alpha}(\vec{M},\theta)$
holds. Its LHS converges locally uniformly to $\theta$.
\item $\forall m, \exists \delta \,> 0$ s.t. 
$\underline{\beta}(\vec{M},\theta,\epsilon) \ge
\underline{\beta}_m(\vec{M},\theta,\epsilon),$ for $\delta \,> 
\forall \epsilon \,> 0$.
\end{itemize}
Similarly to Theorem 2, we can prove inequality (\ref{a11}) 
under the second strong consistency condition.

Under these preparations, we state a theorem with respect to 
the attainability of the bound $J_{\theta}$.
The following theorem can be regarded as a special case of Theorem 8
of \cite{HM}.
\begin{thm}\Label{T5}
Under Assumptions 1 and 3,
the sequence of estimators $\vec{M^s_{\delta}}=
\{ M^{s,n}_{\delta} \}_{n=1}^{\infty}$ 
(defined in the following) satisfies 
the second strong consistency condition and 
the relations
\begin{eqnarray}
\alpha ( \vec{M^s_{\delta}}, \theta)
=\underline{\alpha} ( \vec{M^s_{\delta}}, \theta)
= (1- \delta) \frac{J_{\theta}}{2}. \Label{H5}
\end{eqnarray}
The sequence of estimators $\vec{M^s_{\delta}}$
is independent of the unknown parameter $\theta$.
Every $M^{s,n}_{\delta}$ is an adaptive estimator 
and will be defined in section \ref{s30}.
\end{thm}
Its proof is given in \ref{asd}.

\noindent{\bf [Assumption 3]}:
The following set is compact.
\begin{eqnarray*}
\left\{\left. \left( \Tr \rho_{\theta}\left(
    \frac{L_{\check\theta}}{J_{\check{\theta}}}
    - \frac{\Tr \rho_{\theta}L_{\check\theta}}{J_{\check{\theta}}}
\right)^2\right)^{-1},
\Tr \rho_{\theta}\left(
    \frac{L_{\check\theta}}{J_{\check{\theta}}}
    - \frac{\Tr \rho_{\theta}L_{\check\theta}}{J_{\check{\theta}}}
\right)^2\right| 
\forall \theta, \check{\theta} \in \Theta
\right\} 
\end{eqnarray*}
If the state family is included by  
a bounded closed set consisting of positive definite operators,
Assumption 3 is satisfied.

\noindent
{\bf [Construction of $\vec{M^s_{\delta}}$]}:
We perform a faithful POVM $M_f$ 
(defined in the following) for the first $\delta n$ systems.
Then, the data $(\omega_1, \ldots , \omega_{\delta n})$
obey the probability family $\{ {\rm P}^{M_f}_{\theta}|
\theta \in \Theta \}$.
We denote the maximum likelihood estimator
(MLE) w.r.t.\ the data 
$(\omega_1, \ldots , \omega_{\delta
  n})$ by $\check{\theta}$.
Next, we perform the measurement $E(L_{\check{\theta}})$
defined by the spectral measure 
of $L_{\check{\theta}}$ for other $(1-\delta) n$ systems.
Then, we have data $( \omega_{\delta n +1} , \ldots,
\omega_{n})$.
We decide the final estimated value $T^n_{\check\theta}$ as
\begin{eqnarray*}
\Tr \rho_{T^n_{\check\theta}} L_{\check{\theta}}=
\frac{1}{(1-\delta) n} \sum_{i= \delta n +1}^{n} \omega_i .
\end{eqnarray*}
\endproof
\begin{defi}\rm
A POVM $M$ is called {\it faithful}, if the map
$\rho \in {\cal S}({\cal H}) \mapsto 
{\rm P}^M_{\rho}$ is one-to-one.
\end{defi}
An example of faithful POVM, which is 
a POVM taking values in the set of pure states on ${\cal H}$,
is given by
$M_h( \,d \rho ):= k \rho \nu (\,d \rho )$ ,
where $\nu$ is the invariant (w.r.t.\ the action of $\SU({\cal H})$)
probability measure on the set of pure states on ${\cal H}$.
As another example, if $L_1, \ldots L_{k^2-1}$
is a basis of the space of self-adjoint traceless operators,
a disjoint random combination of PVMs 
$E(L_1), \ldots E(L_{k^2-1})$ is faithful.
Note that a disjoint random combination is defined in section \ref{s1}.

\begin{rem}\Label{rem3}\rm
By dividing $n$ systems into $\sqrt{n}$
and $n- \sqrt{n}$ systems,
Gill and Massar \cite{GillM}
constructed an estimator
which asymptotically attains 
the optimal bound w.r.t.\ MSE,
and Hayashi and Matsumoto \cite{HM98}
constructed a similar estimator
by dividing them into $b_n $ and
$n-b_n$ systems,
where $\lim \frac{b_n}{n}=0$.
However, in our proof,
it is difficult to show the attainability
of the bound (\ref{a11}) in such a division.
Perhaps, there may exist a family in which
such an estimator does not attain
the bound (\ref{a11}).
At least, it is essential in our proof
that the number of the first group $b_n$ satisfy 
$\lim \frac{b_n}{n} \,>0$.

Conversely, as is mentioned in Theorems \ref{4-26} and
\ref{T6},
by dividing $n$ systems into $\sqrt{n}$
and $n- \sqrt{n}$ systems,
we can construct an estimator
attaining the bound (\ref{a3}) at one point.
\end{rem}

We must use quantum correlations 
in the quantum apparatus to achieve the bound
$\frac{\tilde{J}_{\theta}}{2}$.
The following theorem can be easily extended to the multi-parameter
case.
\begin{thm}\Label{T6}
We assume Assumption 1 and that
$D(\rho_{\theta'} \| \rho_{\theta_1}) \,< \infty$ for 
$\forall \theta_1, \forall \theta'\in \Theta$.
Then, for any $\theta_1 \in \Theta$,
the sequence of estimators $\vec{M^w_{\theta_1}}=
\{ M^{w,n}_{\theta_1} \}_{n=1}^{\infty}$ 
satisfies 
the weak consistency condition (\ref{a1}), and 
the equations
\begin{eqnarray}
\underline{\beta}( \vec{M^w_{\theta_1}}, \theta_1, \epsilon )
&=&\beta ( \vec{M^w_{\theta_1}}, \theta_1, \epsilon )
= \inf_{\theta' \in \Theta}
\{ D( \rho_{\theta'} \| \rho_{\theta_1}) | |\theta_1 - \theta' |\,> \epsilon
\}, \Label{G26} \\
\underline{\alpha}(\vec{M^w_{\theta_1}}, \theta_1) &=&
\alpha(\vec{M^w_{\theta_1}}, \theta_1) =
\frac{\tilde{J}_{\theta_1}}{2}. \Label{G27}
\end{eqnarray}
The sequence of estimators $\vec{M^w_{\theta_1}}$
depends on the unknown parameter $\theta_1$
but not on $\epsilon \,> 0$.
\end{thm}
Its proof is given in \ref{ase}.
In the following construction,
$M^{w,n}_{\theta_1}$ is constructed from 
the PVM $E_{\theta_1}^n$, which
is defined from a group-theoretical viewpoint in 
Definition \ref{De2} in \ref{s32}.

\noindent{\bf [Construction of $M^{w,n}_{\theta_1}$]}:
We divide the $n$ systems into two groups.
We perform a faithful POVM $M_f$ for the first group of 
$\sqrt{n}$ systems.
Then, the data $(\omega_1, \ldots , \omega_{\sqrt{n}})$
obey the probability ${\rm P}^{M_f}_{\theta}$.
We let $\check\theta$ be the MLE of the data 
$(\omega_1, \ldots , \omega_{\sqrt{n}})$ under
the probability family
$\{ {\rm P}^{M_f}_{\theta}|\theta \in \Theta \}$.
Next, we perform the correlational PVM 
$E_{\theta_1}^{n- \sqrt{n}}$
for the composite system which consists of 
the other group of $n- \sqrt{n}$ systems.
Then, the data $\omega$ obeys the probability
${\rm P}^{E_{\theta_1}^{n- \sqrt{n}}}_{\theta}$.
If 
$e^{n (1- \delta_{n-\sqrt{n}})D( \rho_{\check{\theta}}\| \rho_{\theta_1})}
{\rm P}^{E_{\theta_1}^{n- \sqrt{n}}}_{\theta_1}
(\omega) \ge 
{\rm P}^{E_{\theta_1}^{n- \sqrt{n}}}
_{\check{\theta}}(\omega)$,
the estimated value $T_n$ is decided to be $\theta_1$,
where $\delta_n := \frac{1}{n^{\frac{1}{5}}}$.
If not, $T_n$ is decided to be $\check{\theta}$.
\endproof

The following lemma proven in \ref{s6}
 plays an important role in the proof of
Theorem \ref{T6}.
\begin{lem}\Label{L8}
For three parameters $\theta_0, \theta_1$ and $\theta_2$
and
$\delta \,> 0$,
the inequalities
\begin{eqnarray}
\fl && {\rm P}_{\theta_0 }^{E^n_{\theta_1}}
\left\{
- \frac{1}{n} \log {\rm P}_{\theta_2}^{E^n_{\theta_1}}(\omega)
+ \Tr \rho_{\theta_0} \log \rho_{\theta_2}
\ge \delta
\right\} \nonumber \\
\fl&\le &
\exp - n \left(
\sup_{0 \le t \le 1}
( \delta - \Tr \rho_{\theta_0}\log \rho_{\theta_2}) t - t \frac{(k+1)\log(n+1)}{n} - \log
\Tr \rho_{\theta_0} {\rho_{\theta_2}}^{-t} \right) \Label{D1} \\
\fl&&{\rm P}_{\theta_0}^{E^n_{\theta_1}}
\left\{
\frac{1}{n} \log {\rm P}_{\theta_1}^{E^n_{\theta_1}}(\omega)
- \Tr \rho_{\theta_0} \log \rho_{\theta_1}
\ge \delta
\right\}\nonumber \\
\fl&\le &
\exp - n \left(
\sup_{0 \le t }
( \delta +\Tr \rho_{\theta_0}\log \rho_{\theta_1}) t - \log \Tr
\rho_{\theta_0} \rho_{\theta_1}^{t} \right) \Label{D2} 
\end{eqnarray}
hold.
\end{lem}
We obtain the following theorem as a 
review of the above discussion.
\begin{thm} \Label{Thm11} From Theorems \ref{T1}, \ref{T2}
and \ref{T5} and Lemma \ref{L3},
we have the equations
\begin{eqnarray}
\sup_{\vec{M}: \hbox{ WC } }
\limsup_{\epsilon \to 0} \frac{1}{\epsilon^2}
\underline{\beta}(\vec{M},\theta,\epsilon) 
&=&
\sup_{\vec{M}: \hbox{ WC } }
\liminf_{\epsilon \to 0} \frac{1}{\epsilon^2}
\underline{\beta}(\vec{M},\theta,\epsilon) 
=\frac{\tilde{J}_{\theta}}{2} 
\Label{a20} \\
 \sup_{\vec{M}: \hbox{ SC at }\theta }
\liminf_{\epsilon \to 0} \frac{1}{\epsilon^2}
\underline{\beta}(\vec{M},\theta,\epsilon) 
&=&\frac{J_{\theta}}{2} 
\Label{G15}
\end{eqnarray}
as an operational comparison of $\tilde{J}_{\theta}$ and $J_{\theta}$
under Assumptions 1, 2 and 3.
We can replace $\beta(\vec{M},\theta,\epsilon) $
with $\underline{\beta}(\vec{M},\theta,\epsilon) $
in equations (\ref{a20}).
\end{thm}
We can also prove (\ref{G12}) as a consequence of
equations (\ref{a20}) and (\ref{G15}).

\section{Adaptive estimators}\Label{s30}
In this section, we assume that
the dimension of the Hilbert space ${\cal H}$
is finite.
We consider estimators
whose POVM is adaptively chosen from the data.
We choose the $l$-th POVM $M_l(\vec{\omega}_{l-1})$ 
on ${\cal H}$ from $l-1$ data $\vec{\omega}_{l-1}:=
(\omega_1, \ldots, \omega_{l-1})$.
Its POVM $M^n$ is described by
\begin{eqnarray}
M^n(\vec{\omega}_n):=
M_1(\omega_1)\otimes M_2(\vec{\omega_1};\omega_2)
\otimes \cdots \otimes M_n(\vec{\omega}_{n-1};\omega_n).
\Label{J1}
\end{eqnarray}
In this setting, the estimator
is written as the pair ${\cal E}_n= (M^n,T_n)$ of
the POVM $M^n$ satisfying (\ref{J1}) and 
the function $T_n: \Omega^n \mapsto \Theta$.
Such an estimator ${\cal E}_n$ is called
an adaptive estimator.
As a larger class of POVMs,
the separable POVM is well known.
A POVM $M^n$ on ${\cal H}^{\otimes n}$ is called separable
if it is written as
\begin{eqnarray*}
M^n=\{ M_1(\omega) \otimes \cdots \otimes M_n(\omega)\}_{\omega \in \Omega}
\end{eqnarray*}
on ${\cal H}^{\otimes n}$, where $M_i(\omega)$ 
is a positive semi-definite operator on ${\cal H}$.
For any separable estimator $(M^n,T_n)$,
the relations
\begin{eqnarray}
D^{M^n}(\theta\|\theta') &=
\sum_{\omega \in \Omega}
\prod_{l'=1}^n
\Tr \rho_\theta M_{l'} (\omega)
\log 
\frac{\prod_{l=1}^n
\Tr \rho_\theta M_l (\omega)}
{\prod_{l=1}^n
\Tr \rho_{\theta'} M_l (\omega)} \nonumber \\
&= \sum_{\omega \in \Omega}
\prod_{l'=1}^n
\Tr \rho_\theta M_{l'} (\omega)
\sum_{l=1}^n
\log 
\frac{\Tr \rho_\theta M_{l} (\omega)}
{\Tr \rho_{\theta'} M_{l} (\omega)} \nonumber \\
&= \sum_{l=1}^n
\sum_{\omega \in \Omega}
a_{\theta,l} (\omega)\Tr \rho_\theta M_l (\omega)
\log 
\frac{a_{\theta,l}(\omega) \Tr \rho_\theta M_{l} (\omega)}
{a_{\theta,l}(\omega) \Tr \rho_{\theta'} M_{l} (\omega)} \nonumber \\
&=\sum_{l=1}^n
D^{M_{\theta,l}}(\theta\|\theta')
\le n \sup_{M: \hbox{POVM on }{\cal H}}
D^M (\theta\|\theta') \Label{H55}
\end{eqnarray}
hold, where the POVM $M_{\theta,l}$ on ${\cal H}$ is 
defined by
\begin{eqnarray*}
 M_{\theta,l}(\omega):= 
a_{\theta,l}(\omega) M_l(\omega), \quad
a_{\theta,l}(\omega):= \left(\prod_{l' \neq l} \Tr \rho_{\theta}
M_{l'}(\omega)\right) .
\end{eqnarray*}
\begin{thm}\Label{T11}
If a sequence of separable estimators
$\vec{M}= \{ {\cal E}_n \}= \{ ( M^n,T_n)\}$
satisfies the weak consistency condition,
the inequalities 
\begin{eqnarray}
\beta( \vec{M},\theta_1, \epsilon)
&\le& \inf_{|\theta- \theta_1| \,> \epsilon}
\sup_{M: \hbox{\rm POVM on }{\cal H}}
D^M(\theta\| \theta_1) \Label{J2}\\
\alpha( \vec{M},\theta_1)
&\le&
\frac{J_{\theta_1}}{2} \Label{J3}
\end{eqnarray}
hold.
\end{thm}
\begin{proof}
Similarly to (\ref{D12}), the monotonicity of quantum relative entropy
yields 
\begin{eqnarray*}
-\frac{\log {\rm P}_{\theta_1}^{M^n}\{ 
|T_n(\vec{\omega}_n)- \theta_1|\,> \epsilon \}}
{n}
\le \frac{D^{M^n}(\theta\| \theta_1)+ h (P_n)}{n P_n},
\end{eqnarray*}
where $P_n:= {\rm P}_{\theta}^{M^n}\{ 
|T_n(\vec{\omega}_n)- \theta_1|\,> \epsilon \}$. From 
the weak consistency, we have $P_n \to 1$.
Thus, we obtain (\ref{J2}) from (\ref{H55}).
Since ${\cal H}$ is finite-dimensional,
the set of extremal points of POVMs is compact.
Therefore,
the convergence $
\lim_{\epsilon \to 0}\frac{1}{\epsilon^2}
D^{M}(\theta_1 + \epsilon \| \theta_1)$ is uniform w.r.t.\
$M$.
This implies that
\begin{eqnarray}
\fl \lim_{\epsilon \to 0}\frac{1}{\epsilon^2}
\sup_{M: \hbox{POVM on }{\cal H}} D^{M}(\theta_1 + \epsilon \| \theta_1)
=
\sup_{M: \hbox{POVM on }{\cal H}}\lim_{\epsilon \to 0}\frac{1}{\epsilon^2}
D^{M}(\theta_1 + \epsilon \| \theta_1)
= \frac{J_{\theta_1}}{2}. \Label{12-13}
\end{eqnarray}
The last equation is derived from (\ref{G14}).
\end{proof}
The preceding theorem holds for any adaptive estimator.
As a simple extension, 
we can define an $m$-adaptive estimator 
that satisfies (\ref{J1})
when every $M_l(\vec{\omega}_{l-1})$ is a POVM on ${\cal H}^{m}$.
As a corollary of Theorem \ref{T11}, we have the following.
\begin{cor}\Label{C5}
If a sequence of $m$-adaptive estimators
$\vec{M}= \{ {\cal E}_n \}= \{ ( M^n,T_n)\}$
satisfies the weak consistency condition,
then the inequalities 
\begin{eqnarray}
\beta( \vec{M},\theta_1, \epsilon)
&\le& \inf_{|\theta- \theta_1| \,> \epsilon}
\sup_{M: \hbox{\rm POVM on }{\cal H}^{\otimes m}}
\frac{1}{m} D^M(\theta\| \theta_1) \Label{J21}\\
\alpha( \vec{M},\theta_1)
&\le&
 \frac{J_{\theta_1}}{2} \Label{J31}
\end{eqnarray}
hold.
\end{cor}
Now, we obtain the equation
\begin{eqnarray}
\lim_{m \to \infty}
\lim_{\epsilon \to 0}
\sup_{\vec{M}:m\hbox{-AWC}}\frac{1}{ \epsilon^2}
\underline{\beta}\vec{M},\theta,\epsilon)=
\frac{J_\theta}{2}\Label{H52}.
\end{eqnarray}
The part of $\ge$ holds because an adaptive estimator 
attaining the bound is constructed in Theorem \ref{T5},
and the part of $\le$ follows from (\ref{J21})
and the equation 
\begin{eqnarray*}
&\lim_{\epsilon \to 0}
\sup_{M: \hbox{POVM on }{\cal H}^{\otimes m}}
\frac{1}{\epsilon^2 m}D^{M}(\theta_1 + \epsilon \| \theta_1) \\
 =&
\sup_{M: \hbox{POVM on }{\cal H}^{\otimes m}}
\lim_{\epsilon \to 0}\frac{1}{\epsilon^2m }
D^{M}(\theta_1 + \epsilon \| \theta_1)
= \frac{J_{\theta_1}}{2},
\end{eqnarray*}
which is proven in a similar manner as (\ref{12-13}).

\section{Difference in order among limits and supremums}
\Label{s9}
Theorem \ref{Thm11}
yields another operational comparison as
\begin{eqnarray}
\sup_{\vec{M}:\hbox{ SC at }\theta}
\liminf_{\epsilon \to 0}
\frac{1}{\epsilon^2}
\underline{\beta}(\vec{M},\theta,\epsilon)
&=& \frac{J_{\theta}}{2} \Label{H1}\\
\lim_{\epsilon \to 0}
\frac{1}{\epsilon^2}
\sup_{\vec{M}:\hbox{ SC at }\theta }
\underline{\beta}(\vec{M},\theta,\epsilon)
&=& \frac{\tilde{J}_{\theta}}{2}\Label{H2}.
\end{eqnarray}
Equation (\ref{H1}) equals (\ref{G15}) and
equation (\ref{H2}) follows from Theorem \ref{Dt1}.
Therefore, the difference between $\frac{J_{\theta}}{2}$ and
$\frac{\tilde{J}_{\theta}}{2}$ can be regarded as the difference in the
order of $\liminf_{\epsilon \to 0}$ and $\sup_{\vec{M}:\hbox{ SC} }$.
This comparison was naively discussed by Nagaoka \cite{Nag,Na1}.
\begin{thm}\Label{Dt1}
We adopt Assumption 1 in Theorem \ref{T5} and 
$D(\rho_{\theta'} \| \rho_{\theta_1}) \,< \infty$ for $\forall \theta' 
\in \Theta$.
For any $\delta \,> 0$,
there exists a sequence $\vec{M}_{\theta_0}^{m,\delta}
= \{ M_{\theta_0}^{m,\delta,n}\}$
of $m$-adaptive estimators
satisfying the strong consistency condition
and the inequality
\begin{eqnarray*}
&& \lim_{n \to \infty}\frac{-1}{nm}
\log {\rm P}_{\theta_0}^{M_{\theta_0}^{m,\delta,n}}
\{ | \hat{\theta}
- \theta_0 | \,> \epsilon \} \\
&\ge&
(1-\delta)
\inf \left\{\left. D (\theta\|\theta_0)\right| 
|\theta - \theta_0|\,> \epsilon\right\}-
\frac{(1-\delta)(k-1)\log (m+1)}{m}.
\end{eqnarray*}
\end{thm}

However, using Theorem \ref{Dt1},
we obtain a stronger equation than (\ref{H2}):
\begin{eqnarray}
\lim_{\epsilon \to 0}
\lim_{m \to \infty}
\sup_{\vec{M}:m\hbox{-ASC at }\theta}\frac{1}{ \epsilon^2}
\beta(\vec{M},\theta,\epsilon)=
\frac{\tilde{J}_\theta}{2} \Label{H51},
\end{eqnarray}
where $m$-ASC at $\theta$ denotes 
$m$-adaptive and is strongly consistent at $\theta$.
This equation is in contrast with (\ref{H52}).
Of course, the part of $\le$ for (\ref{H51}) follows from (\ref{J21}).
The part of $\ge$ for (\ref{H51}) is derived from the above theorem.

The following two lemmas are essential for our proof
of Theorem \ref{Dt1}.
\begin{lem}\Label{L121}
For two parameters $\theta_1$ and $\theta_0$,
the inequality 
\begin{eqnarray}
m D ( \theta_0 \| \theta_1 ) - (k-1)\log (m+1)
\le
D^{E_{\theta_1}^m}( \theta_0 \| \theta_1)
\le
m D ( \theta_0 \| \theta_1 ) \Label{uni}
\end{eqnarray}
holds,
where the PVM $E_{\theta_1}^m$ on ${\cal H}^{\otimes m}$ 
is defined in \ref{s32}.
It is independent of $\theta_0$.
\end{lem}
This lemma was proven by Hayashi \cite{Haya}
and can be regarded as an improvement of Hiai and Petz's 
result \cite{HP}.
However, Hiai and Petz's original version is sufficient 
for our proof 
of Theorem \ref{Dt1}.
For the reader's convenience,
the proof is presented in \ref{s32}.
\begin{lem}\Label{L4thu}
Let $Y$ be a curved exponential family
and $X$ be an exponential family including $Y$.
For a curved exponential family and an exponential family,
see Chap 4 in Amari and Nagaoka \cite{Na} or Barndorff-Nielsen \cite{Barndorff}.
In this setting,  
for n-i.i.d.\ data, the MLE $T_{X,n}^{ML}
(\omega^n)$
for the exponential family $X$ is a sufficient statistic
for the curved exponential family $Y$,
where $\vec{\omega}_n:=(\omega_1, \ldots, \omega_n)$.
Using the map $T : X \to Y$, we can define an estimator $T \circ 
T_{X,n}^{ML}$,
and for an estimator $T_Y$,
there exists a map $T : X \to Y$ such that $T_Y= T \circ 
T_{X,n}^{ML}$.
We can identify a map $T$ from $X$ to $Y$
with a sequence of estimators $T\circ T_{X,n}^{ML}(\vec{\omega}_n)$.
We define the map $T_{\theta_0}: X \to Y$
as
\begin{eqnarray}
T_{\theta_0}:= \arg \min_{\theta \in Y}
\{ D(x \|\theta)| D(\theta\|\theta_0) \le D(x\|\theta_0) \}.
\Label{Dec3}
\end{eqnarray}
When $Y$ is an exponential family (i.e., flat),
$T_{\theta_0}$ coincides with
the projection to $Y$.
Then, 
the sequence of estimators 
corresponding to the map $T_{\theta_0}$ satisfies 
the strong consistency at $\theta_0$
and 
the equation
\begin{eqnarray}
\fl \lim_{n \to \infty}
\frac{-1}{n} \log p^n_{\theta_0}
\{ \| T_{\theta_0}\circ T_{X,n}^{ML}(\vec{\omega}_n) - \theta_0\| 
\,> \epsilon\}
= \inf_{\theta \in Y} \{ D(\theta\| \theta_0)
| \| \theta - \theta_0\| \,> \epsilon \} \Label{H31}
\end{eqnarray}
holds
\end{lem}
\begin{proof}
It is well known that for any subset $X' \subset X$, the equation
\begin{eqnarray}
\lim_{n \to \infty} - \frac{1}{n}
\log p_{\theta_0}^n \{ T_{X,n}^{ML}( \vec{\omega}_n) \in X'\}
= \inf _{x \in X'} D(x \|\theta_0) \Label{Dec5}
\end{eqnarray}
holds. 
For the reader's convenience, we present a proof of (\ref{Dec5})
in \ref{s10}.
Thus, equation (\ref{H31}) follows from 
(\ref{Dec3}) and (\ref{Dec5}).
If $Y$ is an exponential family, then 
the estimator $T_{\theta_0}\circ T_{X,n}^{ML}$ coincides
with the MLE and satisfies the strong consistency.
Otherwise, we choose a neighborhood $U$ of $\theta_0$
so that we can approximate the neighborhood $U$ by 
the tangent space.
The estimator $T_{\theta_0}\circ T_{X,n}^{ML}$ can be approximated
by the MLE and satisfies the strong consistency at $U$.
Thus, it also satisfies 
the strong consistency at $\theta_0$.
\end{proof}
\noindent{\it Proof of Theorem \ref{Dt1}:}\quad
Let $M= \{ M_i \}$ be a faithful POVM defined in section \ref{s52} such that
the number of operators $M_i$ is finite.
For any $m$ and any $\delta \,> 0$, 
we define the POVM $M^m_{\theta_0}$ to be
the disjoint random combination 
of $M \times m$ and $E^m_{\theta_0}$ 
with the ratio $\delta :1-\delta$.
Note that a disjoint random combination is defined in section \ref{s1}. From 
the definition of $M^m_{\theta_0}$,
the inequality
\begin{eqnarray}
(1-\delta) D^{E^m_{\theta_0}}( \theta \| \theta)
\le D^{M^m_{\theta_0}}( \theta \| \theta) \Label{Dec6}
\end{eqnarray}
holds.
Since the map $\theta \mapsto {\rm P}_{\theta}^{M}$ is
one-to-one,
the map $\theta \mapsto {\rm P}_{\theta}^{M^m_{\theta_0}}$ is also
one-to-one.
Since $M$ and $E^m_{\theta_0}$ are finite-resolutions 
of the identity,
the one-parameter family 
$\{ {\rm P}_{\theta}^{M^m_{\theta_0}}|
\theta \in \Theta\}$
is a subset of multi-nominal distributions
$X$, which is an exponential family.
Applying Lemma \ref{L4thu},
we have
\begin{eqnarray*}
& \lim_{n \to \infty}\frac{-1}{nm}
\log {\rm P}_{\theta_0}^{M^m_{\theta_0}\times n}
\{ | T_{\theta_0}\circ T_{X,n}^{ML}( \vec{\omega}_n)
- \theta_0 | \,> \epsilon \}\\
=& \frac{1}{M}
\inf_{\theta \in \Theta}
\{ D^{M^m_{\theta_0}} (\theta \| \theta_0)
 \| | \theta- \theta_0| \,> \epsilon \} \\
\ge& \frac{(1-\delta) }{m}
\inf \left\{\left. D^{E^m_{\theta_0}}(\theta\|\theta_0)\right| 
|\theta - \theta_0|\,> \epsilon\right\}\\
\ge&
(1-\delta)
\inf \left\{\left. D (\theta\|\theta_0)\right| 
|\theta - \theta_0|\,> \epsilon\right\}-
\frac{(1-\delta)(k-1)\log (m+1)}{m},
\end{eqnarray*}
where the first inequality follows from (\ref{Dec6})
and the second inequality follows from (\ref{uni}).
\endproof
\begin{rem}\rm
In the case of the one-parameter equatorial 
spin 1/2 system state family,
the map $\theta \mapsto {\rm P}_{\theta}^{E^m_{\theta_0}}$
is not one-to-one.
Therefore, we must treat not $E^m_{\theta_0}$
but $M^m_{\theta_0}$.
\end{rem}

\section*{Conclusions}
It has been clarified that SLD Fisher information $J_{\theta}$ 
gives the essential 
large deviation bound in the quantum estimation 
and KMB Fisher information $\tilde{J_{\theta}}$ 
gives the large deviation bound
of consistent superefficient estimators.
Since estimators attaining the bound $\frac{\tilde{J_{\theta}}}{2}$ 
are unnatural,
the bound $\frac{J_{\theta}}{2}$
is more important from the viewpoint of quantum estimation
than the bound $\frac{\tilde{J_{\theta}}}{2}$.
On the other hand, as is mentioned in \ref{rem1},
concerning a quantum analogue of information geometry
from the viewpoint of e-connections,
KMB is the most natural among the quantum versions
of Fisher information.
The interpretation of these two facts which
seem to contradict each other, remains a problem.
Similarly, it is a future problem to explain geometrically
the relationship between the change of the orders of limits and
the difference between the two quantum analogues
of Fisher information.
\section*{Acknowledgments}
The author wishes to thank Professor H. Nagaoka
for encouragement and essential advice regarding 
this manuscript.
He also wishes to thank Professor K. Matsumoto for 
useful advice regarding this manuscript.
He is grateful to Professor S. Amari and Professor A. Tomita
and two anonymous referees, whose comments
helped to considerably improve the presentation.

\appendix
\section{Brief review of information-geometrical properties
of $J_{\theta},\tilde{J}_{\theta}$ and $\check{J}_{\theta}$}\Label{rem1}
The quantum analogues of Fisher information
$J_{\theta},\tilde{J}_{\theta}$ and $\check{J}_{\theta}$
are obtained from the 
the inner products
$J_{\rho}$, $\tilde{J}_{\rho}$ and $\check{J}_{\rho}$
on the linear space consisting of
self-adjoint operators:
\begin{eqnarray*}
\tilde{J}_{\rho}(A,B) &:= \Tr A \tilde{L}_B,
\quad \int_0^1 \rho^t \tilde{L}_B \rho^{1-t} \,d t
= B  \\
J_{\rho}(A,B) &:= \Tr A L_B,\quad 
\frac{1}{2}(L_B \rho + \rho L_B )= B \\
\check{J}_{\rho}(A,B) &:=
\Tr A \check{L}_{B}, \quad
B= \rho \check{L}_{B} 
\end{eqnarray*}
in the following way:
\begin{eqnarray*}
J_{\theta}= 
J_{\rho_\theta}\left(\frac{\,d \rho_{\theta}}{\,d \theta},
\frac{\,d \rho_{\theta}}{\,d \theta}\right) \\
\tilde{J}_{\theta}= 
\tilde{J}_{\rho_\theta}
\left(\frac{\,d \rho_{\theta}}{\,d \theta},
\frac{\,d \rho_{\theta}}{\,d \theta}\right) \\
\check{J}_{\theta}= 
\check{J}_{\rho_\theta}
\left(\frac{\,d \rho_{\theta}}{\,d \theta},
\frac{\,d \rho_{\theta}}{\,d \theta}\right).
\end{eqnarray*}
In the multi-dimensional case,
these are regarded as 
metrics as follows.
For example, we can define a metrics
\begin{eqnarray}
\langle \partial_i , \partial_j \rangle = 
J_{\rho_\theta}\left(\frac{\partial \rho_{\theta}}{\partial \theta^i},
\frac{\partial \rho_{\theta}}{\partial \theta^j}\right)
\Label{mat}
\end{eqnarray}
on the tangent space at $\theta$,
and the RHS of (\ref{mat}) is called SLD Fisher matrix.

In quantum setting, any information precessing is described by
a trace-preserving CP (completely positive) map $C: {\cal S}({\cal H} )\to 
{\cal S}({\cal H}')$.
These inner product
satisfy the monotonicity:
\begin{eqnarray*}
J_{\rho_\theta}\left(
\frac{\,d \rho_\theta}{\,d \theta},
\frac{\,d \rho_\theta}{\,d \theta}
\right)
\ge
J_{C(\rho_\theta)}\left(
\frac{\,d C(\rho_\theta)}{\,d \theta},
\frac{\,d C(\rho_\theta)}{\,d \theta}
\right) \\
\tilde{J}_{\rho_\theta}\left(
\frac{\,d \rho_\theta}{\,d \theta},
\frac{\,d \rho_\theta}{\,d \theta}
\right)
\ge
\tilde{J}_{C(\rho_\theta)}\left(
\frac{\,d C(\rho_\theta)}{\,d \theta},
\frac{\,d C(\rho_\theta)}{\,d \theta}
\right) \\
\check{J}_{\rho_\theta}\left(
\frac{\,d \rho_\theta}{\,d \theta},
\frac{\,d \rho_\theta}{\,d \theta}
\right)
\ge
\check{J}_{C(\rho_\theta)}\left(
\frac{\,d C(\rho_\theta)}{\,d \theta},
\frac{\,d C(\rho_\theta)}{\,d \theta}
\right) 
\end{eqnarray*}
for a one-parametric density family
$\{\rho_{\theta} \in {\cal S}({\cal H} )|
\theta \in \Theta \subset \real \}$\cite{Na}.
These inequalities can be regarded as the quantum versions of 
(\ref{j7}).
An inner product satisfying the above
is called a monotone inner product.
According Petz \cite{P10},
the inner product $\check{J}_\rho$ is the 
maximum one among normalized monotone inner products,
and the inner product $J_\rho$ is the 
minimum one.

In the information geometry community,
we usually discuss the torsion.
As is known within this community,
$\alpha$-connection is a generalization 
of $e$-connection.
The torsion of $\alpha$-connection
concerning Fisher inner product 
vanishes in any distribution family\cite{Na}.
In quantum setting,
we can define the $e$-connections
with respect to several quantum Fisher inner products.
One may expect that
in a quantum setting, its torsion
vanishes in any density family.
However, for only the inner product $\tilde{J}_\rho$,
the torsion of $e$-connection vanishes 
in any density family\cite{Na}.
Thus, KMB Fisher information
seems the most natural quantum analogue of
Fisher information,
from an information-geometrical viewpoint.

\section{Proof of (\ref{I12})}\Label{as0} From 
(\ref{j2}), we can calculate as
\begin{eqnarray}
\fl D(\rho_{\theta+\epsilon}\| \rho_{\theta})
&= \Tr \left(
\rho_{\theta+\epsilon}(\log \rho_{\theta+\epsilon}
- \log \rho_{\theta}) \right)
\cong 
\Tr \left(\rho_{\theta}+ \frac{\,d \rho_{\theta}}{\,d \theta}
\epsilon \right)
\left(
\frac{\,d \log \rho_{\theta}}{\,d \theta}\epsilon 
+
\frac{1}{2}\frac{\,d^2 \log \rho_{\theta}}{\,d \theta^2}
\epsilon ^2
\right) \nonumber\\
\fl &=
\Tr \left( \rho_{\theta}\tilde{L_{\theta}}\right) \epsilon
+ 
\left(\Tr \left(\frac{\,d \rho_{\theta}}{\,d \theta}\tilde{L_{\theta}}\right)
+ \frac{1}{2}
\Tr \left(\rho_{\theta} \frac{\,d^2 \log \rho_{\theta}}{\,d \theta^2}\right)
\right)
\epsilon ^2 \Label{j3}.
\end{eqnarray}
Next, we calculate the above coefficients
\begin{eqnarray}
\Tr  \left(\rho_{\theta}\tilde{L_{\theta}} \right)
= \int_0^1 \Tr  \left(\rho_{\theta}^t\tilde{L_{\theta}}
 \rho_{\theta}^{1-t}\right)\,d t
= \Tr \left(\frac{\,d \rho_{\theta}}{\,d \theta}\right)=0  \Label{j1}.
\end{eqnarray}
Using (\ref{j1}) and (\ref{j2}), we have 
\begin{eqnarray}
\fl \Tr \left(\rho_{\theta} \frac{\,d^2 \log \rho_{\theta}}{\,d \theta^2}\right)
& = 
\frac{\,d }{\,d \theta }
\left(\Tr \left(\rho_{\theta}\frac{\,d \log \rho_{\theta}}{\,d \theta}\right)
\right)
-\Tr \left(\frac{\,d \rho_{\theta}}{\,d \theta}
\frac{\,d \log \rho_{\theta}}{\,d \theta} \right) \nonumber \\
& = -\Tr \left(\frac{\,d \rho_{\theta}}{\,d \theta}\tilde{L_{\theta}} \right)
= - \tilde{J_{\theta}} . \Label{j4}
\end{eqnarray}From (\ref{j3}), (\ref{j1}) and (\ref{j4}),
we obtain
\begin{eqnarray*}
D(\rho_{\theta+\epsilon}\| \rho_{\theta})
\cong 
\frac{1}{2} \tilde{J_{\theta}} \epsilon^2 .
\end{eqnarray*}

\section{Proof of Lemma \ref{L2}}\Label{as1}
We define the unitary operator $U_{\epsilon}$ as
\begin{eqnarray*}
b^2(\rho_{\theta}, \rho_{\theta+\epsilon})=
2\left(1- \Tr\left|\sqrt{\rho_{\theta}}\sqrt{\rho_{\theta+ \epsilon}}\right|\right)
= \Tr ( \sqrt{\rho}- \sqrt{\sigma}U_{\epsilon})
( \sqrt{\rho}- \sqrt{\sigma}U_{\epsilon})^*.
\end{eqnarray*}
Letting $W(\epsilon)$ be $\sqrt{\rho_{\theta+\epsilon}}U_{\epsilon}$,
then we have 
\begin{eqnarray*}
b^2(\rho_{\theta}, \rho_{\theta+\epsilon})&=&
\Tr (W(0)- W(\epsilon))(W(0)- W(\epsilon))^* \\
&\cong&
\Tr \left( - \frac{\,d W}{\,d \epsilon}(0)\epsilon\right)  
\left( - \frac{\,d W}{\,d \epsilon}(0)\epsilon\right) ^*
\cong \Tr \frac{\,d W}{\,d \epsilon}(0)\frac{\,d W}{\,d \epsilon}(0)^* \epsilon^2.
\end{eqnarray*}
As is proven in the following discussion, 
the SLD $L$ satisfies
\begin{eqnarray}
\frac{\,d W}{\,d \epsilon}(0) =  \frac{1}{2}L W(0). \Label{19-1}
\end{eqnarray}
Therefore, we have 
\begin{eqnarray*}
b^2(\rho_{\theta}, \rho_{\theta+\epsilon})\cong
\Tr \frac{1}{4}
L W(0)W(0)^* L \epsilon ^2 =
\frac{1}{4}\Tr L^2 \rho_{\theta} \epsilon .
\end{eqnarray*}
We obtain (\ref{a18}).
It is sufficient to show (\ref{19-1}).
\par From the definition of the Bures distance, we have
\begin{eqnarray*}
b^2(\rho_{\theta},\rho_{\theta+\epsilon})
&=&
\min_{U:\hbox{\footnotesize unitary}} \Tr ( \sqrt{\rho_{\theta}}- \sqrt{\rho_{\theta+\epsilon}}U)
( \sqrt{\rho_{\theta}}- \sqrt{\rho_{\theta+\epsilon}}U)^*\\
&=&2- \max_{U:\hbox{\footnotesize unitary}} 
\Tr \sqrt{\rho_{\theta}}
\sqrt{\rho_{\theta+\epsilon}}U^* 
+ U\sqrt{\rho_{\theta+\epsilon}}\sqrt{\rho_{\theta}}\\
&=& 2- \Tr \left|\sqrt{\rho_{\theta}}\sqrt{\rho_{\theta+\epsilon}}\right| 
+ \left|\sqrt{\rho_{\theta+\epsilon}}\sqrt{\rho_{\theta}}\right|\\
&=& 2-  \Tr \left( \sqrt{\rho_{\theta}}\sqrt{\rho_{\theta+\epsilon}}U(\epsilon)^* 
+ U(\epsilon)\sqrt{\rho_{\theta+\epsilon}}\sqrt{\rho_{\theta}} \right),
\end{eqnarray*}
which implies that 
$\sqrt{\rho_{\theta}}\sqrt{\rho_{\theta+\epsilon}}U(\epsilon)^* =
U(\epsilon)\sqrt{\rho_{\theta+\epsilon}}\sqrt{\rho_{\theta}}$.
Therefore, $W(0)W(\epsilon)^*= W(\epsilon)W(0)^*$.
Taking the derivative, we have
\begin{eqnarray*}
W(0) \frac{\,d W}{\,d \epsilon}(0)^* = \frac{\,d W}{\,d \epsilon}(0)W(0)^*,
\end{eqnarray*}
which implies that there exists a self-adjoint operator $L$ 
such that 
\begin{eqnarray*}
\frac{\,d W}{\,d \epsilon}(0)= \frac{1}{2}L
W(0).
\end{eqnarray*}
Since $\rho_{\theta+ \epsilon}= W(\epsilon)W(\epsilon)^*$,
we have
\begin{eqnarray*}
\frac{\,d \rho}{\,d \theta}(\theta)=
\frac{1}{2}\left(L W(0)W(0)^* + W(0)W(0)^*L\right).
\end{eqnarray*}
Thus, the operator $L$ coincides with the SLD.

\section{Proof of (\ref{A7.1})}\Label{apb}
Let $M=\{M_i\}$ be an arbitrary POVM.
We choose the unitary $U$ satisfying
\begin{eqnarray*}
U \sigma^{1/2} \rho^{1/2} =  \sqrt{\rho^{1/2}\sigma\rho^{1/2}}.
\end{eqnarray*}
Using the Schwartz inequality, we have
\begin{eqnarray*}
\fl \sqrt{{\rm P}^{M}_\rho(\omega)}
\sqrt{{\rm P}^{M}_\sigma( \omega)} 
 &=&
\sqrt{ \Tr \left(M_\omega^{1/2} \sigma^{1/2}U^*\right)^*
\left(M_\omega^{1/2} \sigma^{1/2}U^*\right)}
\sqrt{\Tr \left(M_\omega^{1/2} \rho^{1/2}\right)^*
\left(M_\omega^{1/2} \rho^{1/2}\right)} \\
\fl & \ge &
\Tr \left(M_\omega^{1/2} \sigma^{1/2}U^*\right)^*
\left(M_\omega^{1/2} \rho^{1/2}\right)
= \left| \Tr U \sigma^{1/2}M_\omega\rho^{1/2}\right|.
\end{eqnarray*}
Therefore, 
\begin{eqnarray*}
\sum_{\omega} \sqrt{{\rm P}^{M}_\rho(\omega)}
\sqrt{{\rm P}^{M}_\sigma( \omega)}
&\ge&
\sum_{\omega} 
\left|\Tr U \sigma^{1/2}M_\omega\rho^{1/2}\right|
\ge
\left|\sum_{\omega} \Tr U \sigma^{1/2}M_\omega\rho^{1/2}\right|\\
&=&\left|
\Tr U \sigma^{1/2}\rho^{1/2}\right|
= 
\Tr \sqrt{\rho^{1/2}\sigma\rho^{1/2}}.
\end{eqnarray*}
\section{Proof of Lemma \ref{L1}}\Label{as2}
Let $m$ and $\epsilon $ be an arbitrary positive integer
and an arbitrary positive real number,
respectively.
There exists a sufficiently large integer $N$ such that
\begin{eqnarray*}
\frac{1}{n} \log {\rm P}_{\theta}^{M^n}
\left\{ | \hat{\theta} - \theta| \,> \frac{\delta}{m}i\right\} 
\le
- \underline{\beta} \left(\vec{M},\theta,\frac{\delta}{m}i\right) + \epsilon \\
\frac{1}{n} \log {\rm P}_{\theta+\delta}^{M^n}
\left\{ | \hat{\theta} - (\theta+\delta)| \,> \frac{\delta}{m}(m-i)\right\}
\le
- \underline{\beta} \left(\vec{M},\theta+\delta,\frac{\delta}{m}(m-i) \right) 
+ \epsilon
\end{eqnarray*}
for $i = 0,\ldots,m$ and $\forall n \ge N$. From 
the monotonicity (\ref{A7}) and the additivity (\ref{a4}) 
of quantum affinity, we perform the following evaluation:
\begin{eqnarray*}
\fl&& - \frac{n}{8}  I( \rho_{\theta} \| \rho_{\theta+\delta}) 
= - \frac{1}{8}  
I( \rho_{\theta}^{\otimes n} \| \rho_{\theta+\delta}^{\otimes n}) \\
\fl&\le&
\log 
\Biggl(
{\rm P}_{\theta}^{M^n}\left\{
\hat{\theta} \le \theta \right\}^{\frac{1}{2}}
{\rm P}_{\theta+\delta}^{M^n}\left\{
\hat{\theta} \le \theta \right\}^{\frac{1}{2}}
+
{\rm P}_{\theta}^{M^n}\left\{
\theta + \delta \,< \hat{\theta} \right\}^{\frac{1}{2}}
{\rm P}_{\theta+\delta}^{M^n}\left\{
\theta + \delta \,< \hat{\theta}\right\}^{\frac{1}{2}} \\
\fl&& +
\sum_{i=1}^m
{\rm P}_{\theta}^{M^n}\left\{
\theta + \frac{\delta}{m}(i-1) \,< \hat{\theta} 
\le \theta + \frac{\delta}{m}i \right\}^{\frac{1}{2}}
{\rm P}_{\theta+\delta}^{M^n}\left\{
\theta + \frac{\delta}{m}(i-1) \,< \hat{\theta} 
\le \theta + \frac{\delta}{m}i \right\}^{\frac{1}{2}}
\Biggr) \\
\fl&\le&
\log 
\Biggl(
{\rm P}_{\theta+\delta}^{M^n}\left\{
\left|\hat{\theta} - (\theta+ \delta)\right|\ge \delta\right\}^{\frac{1}{2}}
+
{\rm P}_{\theta}^{M^n}\left\{
\left|\hat{\theta} - \theta \right|\,> \delta\right\}^{\frac{1}{2}} \\
\fl&&+
\sum_{i=1}^m
{\rm P}_{\theta}^{M^n}\left\{
\left|\hat{\theta} - \theta \right| \,> \frac{\delta}{m}(i-1)
\right\}^{\frac{1}{2}}
{\rm P}_{\theta+\delta}^{M^n}\left\{
\left|\hat{\theta} - (\theta+ \delta)\right|\ge
\frac{\delta}{m}(m-i) \right\}^{\frac{1}{2}}
\Biggr) \\
\fl&\le&
\log 
\Biggl(
{\rm P}_{\theta+\delta}^{M^n}\left\{
\left|\hat{\theta} - (\theta+ \delta)\right|\,> \frac{\delta}{m}(m-1)
\delta\right\}^{\frac{1}{2}}
+
{\rm P}_{\theta}^{M^n}\left\{
\left|\hat{\theta} - \theta \right|\,> \delta\right\}^{\frac{1}{2}} \\
\fl&&+
\sum_{i=1}^m
{\rm P}_{\theta}^{M^n}\left\{
\left|\hat{\theta} - \theta \right| \,> \frac{\delta}{m}(i-1)
\right\}^{\frac{1}{2}}
{\rm P}_{\theta+\delta}^{M^n}\left\{
\left|\hat{\theta} - (\theta+ \delta)\right|\,>
\frac{\delta}{m}(m-i-1) \right\}^{\frac{1}{2}}
\Biggr) \\
\fl &\le&
\log 
\Biggl(
\exp\left(-\frac{n}{2}\left(\underline{\beta}
\left(\vec{M},\theta,\frac{\delta}{m}(m-1)\right)
- \epsilon\right)\right)
+
\exp\left(-\frac{n}{2}\left(\underline{\beta}
\left(\vec{M},\theta+ \delta,\delta\right)
- \epsilon\right)\right) \\
\fl&&+
\sum_{i=1}^m
\exp\left(-\frac{n}{2}\left(\underline{\beta}
\left(\vec{M},\theta,\frac{\delta}{m}(i-1)\right)
- \epsilon\right)
-\frac{n}{2}\left(\underline{\beta}
\left(\vec{M},\theta+ \delta,\frac{\delta}{m}(m-i-1)\right)
- \epsilon\right)\right)
\Biggr ) \\
\fl&\le&
\log (m+2) \exp
\left(
-\frac{n}{2}
\min_{0 \le i \le m}
\left(\underline{\beta}\left(\vec{M},\theta,\frac{\delta}{m}(i-1)\right)
+
\underline{\beta}\left(\vec{M},\theta+ \delta,\frac{\delta}{m}(m-i-1)\right)
- 2 \epsilon\right)
\right) \\
\fl&=&
\log (m+2) -
\frac{n}{2}\left(
\min_{0 \le i \le m}
\underline{\beta}\left(\vec{M},\theta,\frac{\delta}{m}(i-1)\right)+
\underline{\beta}\left(\vec{M},\theta+ \delta,\frac{\delta}{m}(m-i-1)\right)
- 2 \epsilon\right),
\end{eqnarray*}
where we assume that $\underline{\beta}(\vec{M},\theta,a)=0$ for
any negative real number $a$.
Taking the limit $n \to \infty$ after dividing by $n$,
we have
\begin{eqnarray*}
\fl \frac{1}{8}  I( \rho_{\theta} \| \rho_{\theta+\delta})
\ge \frac{1}{2}
\min_{0 \le i \le m}
\left(\underline{\beta}\left(\vec{M},\theta,\frac{\delta}{m}(i-1)\right)
+
\underline{\beta}\left(\vec{M},\theta+ \delta,\frac{\delta}{m}(m-i-1)\right)
- 2 \epsilon\right).
\end{eqnarray*}
Since $\epsilon \,> 0$ is arbitrary, 
the inequality 
\begin{eqnarray*}
\fl \frac{1}{8}  I( \rho_{\theta} \| \rho_{\theta+\delta})
\ge \frac{1}{2}
\min_{0 \le i \le m}
\left(\underline{\beta}\left(\vec{M},\theta,\frac{\delta}{m}(i-1)\right)
+
\underline{\beta}\left(\vec{M},\theta+ \delta,
\frac{\delta}{m}(m-i-1)\right)
\right)
\end{eqnarray*}
holds.
Taking the limit $m \to \infty$,
we obtain (\ref{a19}).

\section{Unitary evolutions on the boson coherent system}
\Label{s11}
In the system ${\cal H}=L^2(\real)$,
the unitary operator $
U_1(\beta):= \exp(\beta a^* - \beta^* a)$
acts on the coherent state as
\begin{eqnarray*}
U_1(\beta)
| \alpha \rangle =
|\alpha-\beta\rangle,
\end{eqnarray*}
where $\alpha$ and $\beta$ are complex 
numbers and $a$ is the annihilation operator.
Thus, we can verify that
\begin{eqnarray*}
U_1(\beta)
\rho_{\alpha}
U_1(\beta)^*
=\rho_{\alpha- \beta}.
\end{eqnarray*}
Now, we let 
$a_i$ be the annihilation operator on the 
$i$-th system.
The unitary operator
$U_n(\beta) := \prod_{i=1}^n 
\exp(-\beta a_i^* + \beta^* a_i)$
acts on the system 
${\cal H}^{\otimes n}$ as
\begin{eqnarray*}
U_n(\beta)
\rho_{\theta}^{\otimes n} U_n(\beta)^*
=
\rho_{\theta-\beta}^{\otimes n}.
\end{eqnarray*}

In the two-mode system ${\cal H} \otimes {\cal H}$,
the unitary $V_2(t):=
\exp t( -a_2^* a_1+ a_1^* a_2)$ acts as
\begin{eqnarray*}
V_1(t)| \alpha_1 \rangle 
\otimes | \alpha_2\rangle =
| \alpha_1 \cos t + \alpha_2\sin t \rangle
\otimes| -\alpha_1 \sin t+ \alpha_2\cos t \rangle.
\end{eqnarray*}
Thus, we can verify that
\begin{eqnarray*}
V_1(t) \rho_{\theta_1} \otimes \rho_{\theta_2}
V_1(t)^*=
\rho_{\theta_1 \cos t+ \theta_2\sin t  } 
\otimes \rho_{ -\theta_1 \sin t+ \theta_2\cos t }.
\end{eqnarray*}
Therefore,
the unitary 
$V_n:=\prod_{i=1}^n \exp t_i( - a_i^* a_1+ a_1^* a_i)$
satisfies 
\begin{eqnarray*}
V_n \rho_{\theta}^{\otimes n} V_n^*
=
\rho_{\sqrt{n}\theta}\otimes \rho_0^{\otimes (n-1)},
\end{eqnarray*}
where
$\cos t_i = \sqrt{\frac{i-1}{i}},
\sin t_i = \sqrt{\frac{1}{i}}$.

\section{Proof of Proposition \ref{P1}}\Label{asc}
For a proof of Proposition \ref{P1},
we need the following lemma.
\begin{lem}\Label{L12}
Let $g_n(\omega),f_n(\omega)$ be functions on $\Omega$.
Assume that
the functions $\beta_1( \omega) := \lim_{n \to \infty}
\frac{-1}{n} \log f_n(\omega)$ and $
\beta_2(\omega):= \lim_{n \to \infty}
\frac{-1}{n} \log g_n(\omega)$ are continuous.
If the inequality $g_n(\omega) \le 1 $ holds
for any element $\omega \in \Omega$ and any positive integer $n$,
and if there exists a subset $K \subset \Omega$ such that
\begin{eqnarray*}
\lim_{n \to \infty}\frac{-1}{n}
\log \left( \int_{K} f_n (\omega) \,d \omega \right)
\,> \min_{\omega \in \Omega} 
\left(\beta_1( \omega)+ \beta_2( \omega)\right),
\end{eqnarray*}
the relation
\begin{eqnarray*}
\lim_{n \to \infty} \frac{-1}{n} \log \left(
\int_{\Omega} f_n (\omega) g_n(\omega) \,d \omega\right)
=  \min_{\omega \in \Omega} 
\left(\beta_1( \omega)+ \beta_2( \omega)\right)
\end{eqnarray*}
holds.
\end{lem}
Similarly to Lemma \ref{L1}, Lemma \ref{L12} is proven.

Now, we will prove Proposition \ref{P1}.
From the definition of $M^{w,n}$
and the equation
$ \rho_0 = \frac{1}{\overline{N}+1} \sum_k \left(\frac{\overline{N}}{\overline{N}+1}\right)^k | k
\rangle \langle k |$, we have 
\begin{eqnarray*}
\log {\rm P}_{0}^{M^{s,n}}
\{ T_n \,> \epsilon \}
= \log \sum_{k \,> n \epsilon^2}
\left(\frac{\overline{N}}{\overline{N}+1}\right)^k 
= \log \left(\frac{\overline{N}}{\overline{N}+1}\right)^{[ n\epsilon^2]} ,
\end{eqnarray*}
where $[~]$ is a Gauss notation.
Therefore, we obtain
\begin{eqnarray*}
\beta(\vec{M^w}, 0, \epsilon)= \epsilon^2 \log \left( 1+ \frac{1}{\overline{N}}\right),
\end{eqnarray*}
which implies (\ref{D37}).

Next, we prove the strong consistency condition and (\ref{D39}).
We perform the following calculation:
\begin{eqnarray}
{\rm P}^{M^{w,n}}_{\theta}
\{ T_n - \theta \,> \epsilon\} 
&=& \sum_{k \,> (\theta+\epsilon)^2 n }
\langle k| \int_{\complex} \frac{1}{\pi \overline{N}}| \alpha \rangle \langle
\alpha | e^{-\frac{|\alpha - \sqrt{n}\theta|^2}{\overline{N}}} \,d^2 \alpha
| k \rangle \nonumber\\
&= &
\int_{\complex} \frac{\sqrt{n}}{\pi \overline{N}}
e^{- n \frac{|\alpha - \theta|^2}{\overline{N}}}
\sum_{k \,> (\theta- \epsilon)^2 n}
\frac{(n |\alpha|^2)^k}{k !}e^{-n |\alpha|^2} \,d^2 \alpha . \Label{D32}
\end{eqnarray}
The equation
\begin{eqnarray}
\lim_{n \to \infty}
\frac{-1}{n}\log \frac{\sqrt{n}}{\pi \overline{N}}
e^{- n \frac{|\alpha - \theta|^2}{\overline{N}}} = \frac{|\alpha- \theta|^2}{\overline{N}}
\Label{G45}
\end{eqnarray}
holds.
Also, as is proven in the latter, the equations
\begin{eqnarray}
&& \lim_{n \to \infty}
\frac{-1}{n}\log \left(
\sum_{k \,> (\theta+ \epsilon)^2 n}
\frac{(n |\alpha|^2)^k}{k !}e^{-n |\alpha|^2}\right)\nonumber\\
&=&
\left( (\theta +\epsilon)^2 \log 
\frac{(\theta+\epsilon)^2}{|\alpha|^2}
+ |\alpha|^2 - (\theta+\epsilon)^2\right)
1( (\theta + \epsilon)^2 - |\alpha|^2) \Label{D31}\\
&& \lim_{n \to \infty}
\frac{-1}{n}\log \left(
\sum_{k \,< (\theta- \epsilon)^2 n}
\frac{(n |\alpha|^2)^k}{k !}e^{-n |\alpha|^2}\right)\nonumber \\
&=&
\left( (\theta -\epsilon)^2 \log 
\frac{(\theta-\epsilon)^2}{|\alpha|^2}
+ |\alpha|^2 - (\theta-\epsilon)^2\right)
1( - (\theta - \epsilon)^2 + |\alpha|^2) 
\Label{D31.2}
\end{eqnarray}
hold, 
where $1(x)$ is defined as
\begin{eqnarray*}
1(x) = \left \{
\begin{array}{cc}
1 & x \ge 0 \\
0 & x \,< 0 .
\end{array}
\right.
\end{eqnarray*}
For any $\delta \,> 0$, there
exists a real number $K$ such that
\begin{eqnarray*}
\lim_{n \to \infty}
- \frac{1}{n}
\log \left( \int_{|\alpha|\,> K}
\frac{\sqrt{n}}{\pi \overline{N}}
\exp
\left(- n \frac{|\alpha - \theta|^2}{\overline{N}}
\right)
\,d x \right)
= \frac{K- \theta}{\overline{N}} \,> \delta.
\end{eqnarray*}
Now, we can apply Lemma \ref{L12} to (\ref{D32}). From 
(\ref{G45}) and (\ref{D31}), the relations 
\begin{eqnarray*}
\fl && \lim_{n \to \infty}
\frac{-1}{n} \log
{\rm P}^{M^{w,n}}_{\theta}
\{ T_n - \theta \,> \epsilon\}  \\
\fl &=&
\min_{ \alpha \in \complex}
\left( \frac{|\alpha- \theta|^2}{\overline{N}}+
\left( (\theta +\epsilon)^2 \log 
\frac{(\theta+\epsilon)^2}{|\alpha|^2}
+ |\alpha|^2 - (\theta+\epsilon)^2\right)
1( (\theta + \epsilon)^2 - |\alpha|^2)  \right) \\
\fl &=&
\min_{ \alpha \in \real}
\left( \frac{|\alpha- \theta|^2}{\overline{N}}+
\left( (\theta +\epsilon)^2 \log 
\frac{(\theta+\epsilon)^2}{|\alpha|^2}
+ |\alpha|^2 - (\theta+\epsilon)^2\right)
1( (\theta + \epsilon)^2 - |\alpha|^2)  \right)\\
\fl &=&
\min_{ s \in \real}
\left( \frac{s^2}{\overline{N}}+
\left( (\theta +\epsilon)^2 \log 
\frac{(\theta+\epsilon)^2}{(\theta - s)^2}
+ (\theta - s)^2 - (\theta+\epsilon)^2\right)
1( (\theta + \epsilon)^2 - (\theta - s)^2)  \right)
\end{eqnarray*}
hold.
If $\epsilon $ is sufficiently small for $\theta$,
we have the following approximation:
\begin{eqnarray*}
\fl \lim_{n \to \infty}
\frac{-1}{n} \log
{\rm P}^{M^{w,n}}_{\theta}
\{ T_n - \theta \,> \epsilon\}  
\cong
\min_s \frac{1+2\overline{N}}{\overline{N}}
\left( s- \frac{2\overline{N}}{1+2\overline{N}}\epsilon\right)^2 
+ \frac{\epsilon^2}{\overline{N} + \frac{1}{2}}.
\end{eqnarray*}
Thus, 
\begin{eqnarray}
\lim_{\epsilon \to 0}\lim_{n \to \infty}
\frac{-1}{n\epsilon^2} \log
{\rm P}^{M^{w,n}}_{\theta}
\{ T_n - \theta \,> \epsilon\}  
=\frac{1}{\overline{N} + \frac{1}{2}}.\Label{D33}
\end{eqnarray}
The second convergence of the LHS of (\ref{D33}) 
is uniform in a sufficiently small neighborhood $U_{\theta_0}$ of
arbitrary $\theta_0 \in \real^+ \setminus \{ 0\}$.

Similarly to (\ref{D33}),
from (\ref{D31.2}), we can prove 
\begin{eqnarray}
\lim_{\epsilon \to 0}\lim_{n \to \infty}
\frac{-1}{n\epsilon^2} \log
{\rm P}^{M^{w,n}}_{\theta}
\{ T_n - \theta \,< - \epsilon\}  
=\frac{1}{\overline{N} + \frac{1}{2}}.\Label{D33.1}
\end{eqnarray}
Also, the second convergence of the LHS of (\ref{D33.1}) 
is uniform at a sufficiently small neighborhood $U_{\theta_0}$ of 
arbitrary $\theta_0 \in \real^+ \setminus \{ 0\} $.
Thus, (\ref{D39}) and the strong consistency condition are
proven.

Next, we prove (\ref{D31}) and (\ref{D31.2}).
Using the Stirling formula, we have
\begin{eqnarray}
\fl \lim_{n \to \infty}
\frac{-1}{n} \log 
\frac{(n |\alpha|^2)^{[\delta n]}}{[\delta n] !}e^{-n |\alpha|^2}
=
\left( \delta \log 
\frac{\delta}{|\alpha|^2}
+ |\alpha| - \delta^2\right)
1( \delta - |\alpha|^2) . \Label{D31.1}
\end{eqnarray}
Since the relations
\begin{eqnarray*}
\fl \frac{(n |\alpha|^2)^{([(\theta- \epsilon)^2 n]-1)}}
{([(\theta- \epsilon)^2 n]-1) !}e^{-n |\alpha|^2}
\le 
\sum_{k \,< (\theta- \epsilon)^2 n}
\frac{(n |\alpha|^2)^k}{k !}e^{-n |\alpha|^2}
\le 
[(\theta- \epsilon)^2 n]
\frac{(n |\alpha|^2)^{([(\theta- \epsilon)^2 n]-1)}}
{([(\theta- \epsilon)^2 n]-1) !}e^{-n |\alpha|^2}
\end{eqnarray*}
hold, 
(\ref{D31.2}) follows from (\ref{D31.1}).
If $(\theta+ \epsilon )^2 \le |\alpha|^2$, the equation
\begin{eqnarray}
\lim_{n \to \infty}
\frac{-1}{n}\log 
\sum_{k \,> (\theta+ \epsilon)^2 n}
\frac{(n |\alpha|^2)^k}{k !}e^{-n |\alpha|^2}
= 0  \Label{G44}
\end{eqnarray}
holds. It implies (\ref{D31}) in the case of
$(\theta+ \epsilon )^2 \le |\alpha|^2$.

Next we prove (\ref{D31}) in the case of
$(\theta+ \epsilon )^2 \,> |\alpha|^2$.
In this case, we have 
\begin{eqnarray}
\sum_{L n \,> 
k \,> (\theta+ \epsilon)^2 n}
\frac{(n |\alpha|^2)^k}{k !}e^{-n |\alpha|^2}
\le n( L-(\theta+\epsilon)^2) 
\frac{(n |\alpha|^2)^{[ (\theta+ \epsilon)^2 n]}}
{  [(\theta+ \epsilon)^2 n]!}e^{-n |\alpha|^2}  \Label{G42}
\end{eqnarray}
because $\left(\frac{(n |\alpha|^2)^k}{k !}e^{-n |\alpha|^2}\right)
/ \left(\frac{(n |\alpha|^2)^{(k+1)}}{(k+1) !}e^{-n |\alpha|^2}\right)
= \frac{k+1}{n | \alpha|^2}$.
If $L$ and $N$ are sufficiently large for $|\alpha|^2$,
we have
\begin{eqnarray}
\sum_{k \ge L n}
\frac{(n |\alpha|^2)^k}{k !}e^{-n |\alpha|^2}
\le
\sum_{k \ge L n} e^{-k}
= \frac{e^{- nL}}{1- e^{-1}}\Label{G43}
\end{eqnarray}
because 
(\ref{D31.1}) implies that
\begin{eqnarray*}
\frac{(n |\alpha|^2)^{[\delta n]}}{[\delta n] !}e^{-n |\alpha|^2}
\le e^{-[\delta n] }, \quad \forall \delta \ge L, \forall n \ge N.
\end{eqnarray*}
Since the relations
\begin{eqnarray*}
 \frac{(n |\alpha|^2)^{[ (\theta+ \epsilon)^2 n]}}
{  [(\theta+ \epsilon)^2 n]!}e^{-n |\alpha|^2}  
&\le&
\sum_{k \,> (\theta+ \epsilon)^2 n}
\frac{(n |\alpha|^2)^k}{k !}e^{-n |\alpha|^2} \\
&\le &
n( L-(\theta+\epsilon)^2) 
\frac{(n |\alpha|^2)^{[ (\theta+ \epsilon)^2 n]}}
{  [(\theta+ \epsilon)^2 n]!}e^{-n |\alpha|^2}  
+ \frac{e^{- nL}}{1- e^{-1}}
\end{eqnarray*}
hold,
we have
\begin{eqnarray*}
&& \left( (\theta +\epsilon)^2 \log 
\frac{(\theta+\epsilon)^2}{|\alpha|^2}
+ |\alpha|^2 - (\theta+\epsilon)^2\right) \\
&\ge &
\lim_{n \to \infty}
\frac{-1}{n}\log \left(
\sum_{k \,> (\theta+ \epsilon)^2 n}
\frac{(n |\alpha|^2)^k}{k !}e^{-n |\alpha|^2}\right) \\
&\ge& \min \left\{
\left( (\theta +\epsilon)^2 \log 
\frac{(\theta+\epsilon)^2}{|\alpha|^2}
+ |\alpha|^2 - (\theta+\epsilon)^2\right),
L \right\}.
\end{eqnarray*}
If we let $L$ be a sufficiently large real number, we have (\ref{D31}).

\section{Proof of Theorem \ref{T5}}\Label{asd}
In this proof, we use the function $\phi_{\theta,\check{\theta}}(s)$
defined in (\ref{14-2}).
First, we prove the following four facts.
\begin{description}
\item[(i)]
The faithful POVM $M_f$ satisfies the inequalities 
\begin{eqnarray*}
\beta(\vec{M_f}, \theta, \epsilon) 
\,> 0, \quad
\alpha( \vec{M_f}, \theta) \,> 0 .
\end{eqnarray*}
\item[(ii)] The relation
\begin{eqnarray*}
\lim_{\check{\theta}\to \theta}
\left( \Tr \rho_{\theta}\left(
\frac{L_{\check{\theta}}}{J_{\check{\theta}}}-
\frac{\Tr \rho_{\theta} L_{\check{\theta}}}{J_{\check{\theta}}}
\right)^2\right)^{-1}
= J_{\theta}, \quad
\forall \theta \in \Theta
\end{eqnarray*}
holds.
\item[(iii)]
The equation 
\begin{eqnarray}
\lim_{s\to 0}
\frac{\phi_{\theta,\check{\theta}}(s)-1}{s^2}=
\frac{1}{2}\Tr \rho_{\theta}
\left(
\frac{L_{\check{\theta}}}{J_{\check{\theta}}}-
\frac{\Tr \rho_{\theta} L_{\check{\theta}}}{J_{\check{\theta}}}
\right)^2 \Label{H8.3}
\end{eqnarray}
holds.
The LHS converges uniformly w.r.t.\
$\theta,\check{\theta}$. 
\item[(iv)]
For any real number $\delta_2 \,> 0$, there exists
a sufficiently small real number $\epsilon \,> 0$ such that
if $| \Tr \rho_{\theta} L_{\check{\theta}} - \Tr \rho_{\theta'}
L_{\check{\theta}} | \le \epsilon ( 1- \delta_2)$
and $| \check{\theta} - \theta|\,< \sqrt{\epsilon }$,
then $| \theta' - \theta| \,< \epsilon$.
\end{description} Fact (i) is easily proven from the 
definition of $M_f$. Fact (iii) is proven by the relation
\begin{eqnarray*}
\sup_{\check{\theta},\theta }\left\| 
\frac{L_{\check{\theta}}}{J_{\check{\theta}}}-
\frac{\Tr \rho_{\theta} L_{\check{\theta}}}{J_{\check{\theta}}}
\right\|\,< \infty.
\end{eqnarray*} Fact (ii) is, also, proven by
the relations
\begin{eqnarray*}
\Tr \rho_{\theta}
\left(
\frac{L_{\check{\theta}}}{J_{\check{\theta}}}-
\frac{\Tr \rho_{\theta} L_{\check{\theta}}}{J_{\check{\theta}}}
\right)^2 =
\frac{\Tr \rho_{\theta}\left(L_{\check{\theta}}^2\right)}
{J_{\check{\theta}}^2} -
\frac{\left(\Tr \rho_{\theta}L_{\check{\theta}}\right)^2}
{J_{\check{\theta}}^2} 
\to J_{\theta}^{-1} \hbox{ as } \check\theta \to \theta.
\end{eqnarray*} Fact (iv) follows from 
the relation 
\begin{eqnarray*}
\frac{\partial \Tr \rho_{\theta}L_{\check{\theta}}}
{\partial \theta} \to 1 \hbox{ as } 
\check{\theta} \to \theta ,
\end{eqnarray*}
which follows from fact (i).

Next, we prove the theorem from the preceding four facts.
The inequality 
\begin{eqnarray}
\fl && {\rm P}_{\theta}^{M_{\delta}^{s,n}}
\{\hat{\theta} \notin U_{\theta,\epsilon} \}\nonumber \\
\fl &\le&
{\rm P}_{\theta}^{M_f \times \delta n }
\{ \hat{ \theta} \in U_{\theta,\sqrt{\epsilon}} \}
\sup_{\check{\theta} \in U_{\theta,\sqrt{\epsilon}}}
{\rm P}_{\theta}^{L_{\check{\theta}} \times (1-\delta)n}
\{ \hat{\theta} \notin  U_{\theta,\epsilon} \}
+
{\rm P}_{\theta}^{M_f \times \delta n }
\{ \hat{\theta} \notin U_{\theta,\sqrt{\epsilon}} \}\Label{H4}
\end{eqnarray}
holds.
As is proven in the latter,
the inequality
\begin{eqnarray}
\fl &&\liminf_{n \to \infty}
-\frac{1}{n}\log\sup_{\check\theta \in U_{\theta,\sqrt{\epsilon}}}
{\rm P}_{\theta}^{L_{\check\theta}\times (1-\delta)n} \left\{ 
T_{\check\theta}^n \notin U_{\theta,\epsilon}\right\} \nonumber \\
\fl &\ge& 
(1-\delta) g \left( 
\epsilon^2 (1-\delta_2)^2
\frac{1}{2} \left(\Tr \rho_{\theta}
\left(  \frac{L_{\check\theta}}{J_{\check\theta}}
- \frac{\Tr \rho_{\theta}L_{\check\theta}}{J_{\check\theta}}\right)^2
\right)^{-1},
\frac{\epsilon^2(1-\delta_2)^2}{2}\delta \right) \Label{B3}
\end{eqnarray}
holds, where the function $g(x,y)$ is defined as
$g(x,y):= x- \log(1+\frac{x}{2}+y)$.
Therefore, we have
\begin{eqnarray}
\fl && \underline{\beta}( \vec{M}^s_\delta, \theta, \epsilon)
= \liminf_{n \to \infty} - \frac{1}{n}
\log {\rm P}_{\theta}^{M_{\delta}^{s,n}}
\{ \hat{\theta} \notin U_{\theta,\sqrt{\epsilon}} \}\nonumber \\
\fl &\ge& 
\min \Biggl\{ (1-\delta)h 
\left( 
\epsilon^2 (1-\delta_2)^2
\frac{1}{2} \left(\Tr \rho_{\theta}
\left(  \frac{L_{\check\theta}}{J_{\check\theta}}
- \frac{\Tr \rho_{\theta}L_{\check\theta}}{J_{\check\theta}}\right)^2
\right)^{-1},
\frac{\epsilon^2(1-\delta_2)^2}{2}\delta \right),
\nonumber \\
\fl && \quad
c \underline{\beta}( \{ M_f \times \delta n\}, \theta,
\sqrt{\epsilon}) \Biggr\}.
\Label{B4}
\end{eqnarray} From facts (i) and (ii),
the equations
\begin{eqnarray}
\fl &&\lim_{\epsilon \to 0} \frac{1}{\epsilon^2} (\hbox{RHS of (\ref{B4})}) \nonumber \\
\fl &=& \frac{1-\delta}{2}\left(
\lim_{\check\theta \to \theta} 
(1-\delta_1)^2(1-\delta_2)^2
\left(\Tr \rho_{\theta}
\left(  \frac{L_{\check\theta}}{J_{\check\theta}}
- \frac{\Tr \rho_{\theta}L_{\check\theta}}{J_{\check\theta}}\right)^2
\right)^{-1} - (1-\delta_2)^2\delta_3 \right) \nonumber \\
\fl &=& \frac{1-\delta}{2}\left(
(1-\delta_1)^2(1-\delta_2)^2
J_{\theta} - (1-\delta_2)^2\delta_3 \right)
\Label{B10}
\end{eqnarray}
hold. The RHS of (\ref{B10}) converges
locally uniformly w.r.t.\ $\theta$.
Let $\underline{\beta}_m( \vec{M}_\delta^s, \theta,\epsilon)$ be the
RHS of (\ref{B4}) in the case of $\delta_2=\delta_3=
\frac{1}{m}$.
Therefore, we have
\begin{eqnarray*}
\lim_{m \to \infty} \lim_{\epsilon \to 0}\frac{1}{\epsilon^2}
\underline{\beta}_m( \vec{M}_\delta^s, \theta,\epsilon)
= \frac{1-\delta}{2} J_{\theta},
\end{eqnarray*}
which implies that 
\begin{eqnarray*}
\underline{\alpha}( \vec{M}_\delta^s, \theta) \ge \frac{1-\delta}{2} J_{\theta}.
\end{eqnarray*}
If the converse inequality 
\begin{eqnarray}
\alpha( \vec{M}_\delta^s, \theta) \le \frac{1-\delta}{2}
J_{\theta} \Label{B2}
\end{eqnarray}
holds,
we can immediately derive relations (\ref{H5}) 
and show that 
the sequence of estimators $\vec{M}_\delta^s$ satisfies the
second strong consistency condition.

In the following, the relations (\ref{B2}) and (\ref{B3}) are proven.
First, we prove (\ref{B2}).
We can evaluate the probability 
${\rm P}_{\theta}^{M_\delta^{s,n}}\{ \hat\theta \in
U_{\theta,\epsilon}\}$ as
\begin{eqnarray*}
\fl - \log {\rm P}_{\theta}^{M_\delta^{s,n}}\{ \hat\theta \in
U_{\theta,\epsilon}\} 
&=& - \log \int {\rm P}_{\theta}^{M_f \times \delta n}( \,d \check\theta)
{\rm P}_{\theta}^{L_{\check\theta} \times (1-\delta)n}
\{ T_{\check\theta}^n \notin U_{\theta,\epsilon}\} \\
\fl &\le& - \int {\rm P}_{\theta}^{M_f \times \delta n}( \,d \check\theta)
\log \left({\rm P}_{\theta}^{L_{\check\theta} \times (1-\delta)n}
\{ T_{\check\theta}^n \notin U_{\theta,\epsilon}\} \right)\\
\fl &\le & - \int {\rm P}_{\theta}^{M_f \times \delta n}( \,d \check\theta)
\frac{D^{L_{\check{\theta}}\times (1-\delta)n}
(\theta+ \xi  \epsilon\|\theta)+
h({\rm P}_{\theta+ \xi  \epsilon,n}^{L_{\check{\theta}}})}
{{\rm P}_{\theta+ \xi  \epsilon,n}^{L_{\check{\theta}}}},
\end{eqnarray*}
where ${\rm P}_{\theta+ \xi  \epsilon,n}^{L_{\check{\theta}}}:=
{\rm P}_{\theta+ \xi  \epsilon,n}^{L_{\check{\theta}}\times (1-\delta)n}
\{ T_{\check\theta^n} \notin U_{\theta,\epsilon} \}$,
and similarly to (\ref{D12}), we can prove 
the last inequality.
For any  $\delta_4 \, > 0$, we have
\begin{eqnarray*}
\fl &&\limsup_{n \to \infty}
- \frac{1}{n} \log {\rm P}_\theta^{\vec{M}_\delta^s}
\{ T_n \notin U_{\theta,\epsilon}\} \\
\fl &\le& \limsup_{n \to \infty}
\int_\real {\rm P}_{\theta}^{M_f \times \delta n}( \,d \check\theta)
(1-\delta) \min_{\xi = 1- \delta_4 , -(1- \delta_4)}
\frac{(1-\delta) D^{L_{\check{\theta}}}
(\theta+ \xi  \epsilon\|\theta)+
\frac{h({\rm P}_{\theta+ \xi  \epsilon,n}^{L_{\check{\theta}}})}{n}}
{(1-\delta){\rm P}_{\theta+ \xi  \epsilon,n}^{L_{\check{\theta}}}} \\
\fl &=&
(1-\delta) 
\min_{\xi = 1- \delta_4 , -(1- \delta_4)}
D^{L_{\check{\theta}}}
(\theta+ \xi  \epsilon\|\theta) 
= \frac{1-\delta}{2} J_{\theta} .
\end{eqnarray*}
The last equation is derived from Lebesgue's convergence theorem and
the fact that 
the probability ${\rm P}_{\theta+ \xi  \epsilon,n}^{L_{\check{\theta}}}$ 
tends to $1$ uniformly w.r.t.\ $\check\theta$, as
follows from Assumptions 1 and 3.
 
The reason for the applicability of Lebesgue's convergence 
theorem
is given as follows.
Since
${\rm P}_{\theta+ \xi  \epsilon,n}^{L_{\check{\theta}}}$ 
tends to $1$ uniformly w.r.t.\ $\check\theta$,
there exists $N,R \,> 0$ such that
${\rm P}_{\theta+ \xi  \epsilon,n}^{L_{\check{\theta}}} \,>
\frac{1}{R}, \forall \check\theta \in \Theta , n \ge N$.
Thus, we have
\begin{eqnarray*}
\fl \frac{D^{L_{\check{\theta}}\times (1-\delta)n}
(\theta+ \xi  \epsilon\|\theta)+
h({\rm P}_{\theta+ \xi  \epsilon,n}^{L_{\check{\theta}}})}
{{\rm P}_{\theta+ \xi  \epsilon,n}^{L_{\check{\theta}}}}
\le
\frac{R}{1-\delta}\left(
(1-\delta) D( \theta + \epsilon\xi \| \theta) +2\right) \,< \infty .
\end{eqnarray*}
Therefore, we can apply Lebesgue's convergence theorem.
Thus, the relations 
\begin{eqnarray*}
\alpha( \vec{M}_\delta^s, \theta) 
&=& \limsup_{\epsilon \to 0} \limsup_{n \to \infty} 
- \frac{1}{n \epsilon^2} \log {\rm P}_\theta^{\vec{M}_\delta^s}
\{ T_n \notin U_{\theta,\epsilon} \} \\
&\le& (1-\delta) \limsup_{\epsilon \to 0}\frac{1}{\epsilon^2}
\min_{\xi = 1- \delta_4 , -(1- \delta_4)}
D^{L_{\check{\theta}}}
(\theta+ \xi  \epsilon\|\theta)  \\
&=& (1-\delta)(1- \delta_4)^2 
\frac{1}{2} J_{\theta} 
\end{eqnarray*}
hold.
Since $\delta_4 \,> 0$ is arbitrary,
the inequality (\ref{B2}) holds.

Next, we prove the inequality (\ref{B3}).
Assume that $| \check{\theta}- \theta| \le \epsilon$
and define
\begin{eqnarray*}
\Lambda(\xi, \check{\theta}, \theta):= \
\sup_{\eta \in \real } ( \eta \xi - \log
\phi_{\theta,\check{\theta}}(\eta)).
\end{eqnarray*}
Then, the inequalities
\begin{eqnarray}
\fl {\rm P}_{\theta}^{L_{\check{\theta}}\times (1-\delta)n}
\{ \check{\theta}\notin U_{\theta,\epsilon}\}
&\le& {\rm P}_{\theta}^{L_{\check{\theta}}\times (1-\delta)n}
\left\{ \left| \Tr \rho_{\hat{\theta}}L_{\check\theta}- 
\Tr \rho_{\theta}L_{\check\theta}\right| \le
(1-\delta_2)\epsilon \right\} \Label{H6} \\
\fl &\le&
2 \exp \left(
-(1-\delta)n \min\left\{
\Lambda( (1-\delta_2)\epsilon, \check{\theta}, \theta),
\Lambda( - (1-\delta_2)\epsilon, \check{\theta}, \theta)\right\}
\right)\Label{H7}
\end{eqnarray}
hold, where
(\ref{H6}) is derived from fact (iv), and (\ref{H7}) is derived from 
Markov's inequality.
Thus,
\begin{eqnarray}
&&\lim_{n \to \infty}
- \frac{1}{n}\log
\sup_{\check{\theta} \in U_{\theta,\sqrt{\epsilon}}}
{\rm P}_{\theta}^{L_{\check{\theta}}\times (1-\delta)n}
\{ \check{\theta}\notin U_{\theta,\epsilon}\} \nonumber \\
&\ge &
(1-\delta) \inf_{\check{\theta} \in U_{\theta,\sqrt{\epsilon}}}
\min
\left\{
\Lambda( (1-\delta_2)\epsilon, \check{\theta}, \theta),
\Lambda( - (1-\delta_2)\epsilon, \check{\theta}, \theta)\right\}.
\Label{B11}
\end{eqnarray}
We let $\epsilon \,> 0$ be a sufficiently small real number for arbitrary 
$\delta_3 \,>0$
and define $\eta$ by
\begin{eqnarray*}
\eta:= \epsilon ( 1-\delta_2) 
\left( \Tr \rho_{\theta}\left(
    \frac{L_{\check\theta}}{J_{\check{\theta}}}
    - \frac{\Tr \rho_{\theta}L_{\check\theta}}{J_{\check{\theta}}}
\right)^2\right)^{-1}.
\end{eqnarray*}
Then, the inequalities
\begin{eqnarray}
&& \Lambda( \pm (1-\delta_2)\epsilon, \check{\theta}, \theta) \nonumber \\
&\ge& \pm ( 1- \delta_2) \epsilon (\pm \eta)- \log
\phi_{\theta,\check{\theta}} (\pm \eta) \nonumber \\
& \ge &
\epsilon^2 ( 1- \delta)^2 
\left(\Tr \rho_{\theta}\left(
    \frac{L_{\check\theta}}{J_{\check{\theta}}}
    - \frac{\Tr \rho_{\theta}L_{\check\theta}}{J_{\check{\theta}}}
\right)^2\right)^{-1} \nonumber \\
&& - \log \left(
1 + \frac{\epsilon^2( 1- \delta)^2}{2}
\left( 
\left( \Tr \rho_{\theta}\left(
    \frac{L_{\check\theta}}{J_{\check{\theta}}}
    - \frac{\Tr \rho_{\theta}L_{\check\theta}}{J_{\check{\theta}}}
\right)^2\right)^{-1}
+ \delta_3 
\right)\right) \Label{H8}
\end{eqnarray}
hold, where (\ref{H8}) follows from fact (iii).
The uniformity of (\ref{H8.3}) (the fact(iii)) and 
the boundedness of RHS of (\ref{H8.3}) (Assumption 3)
guarantee that the choice of $\epsilon \,> 0$ is independent of $\theta,\check\theta$. From (\ref{B11})
and (\ref{H8}),
we obtain (\ref{B4}) because 
the function $x \mapsto g(x,y)$ where $y,x \ge 0$.
\section{Proof of Theorem \ref{T6}}\Label{ase}
If the true state is $\rho_{\theta_1}$,
the inequalities 
\begin{eqnarray*}
\fl &&{\rm P}_{\theta_1}^{M^{w,n}_{ \theta_1}}
\{ T_n \notin U_{\theta_1,\epsilon}\}\\
\fl &\le&
{\rm P}^{M_f \times \sqrt{n}}_{\theta_1}
\{ \check\theta \notin  U_{\theta_1,\epsilon}\}
\sup_{\check\theta \notin  U_{\theta_1,\epsilon}}
{\rm P}_{\theta_1}^{E_{\theta_1}^{n- \sqrt{n}}}
\left \{
e^{n (1- \delta_{n- \sqrt{n}})D( {\check{\theta}}\| {\theta_1})}
{\rm P}^{E_{\theta_1}^{n- \sqrt{n}}}_{\theta_1}
(\omega) \,<
{\rm P}^{E_{\theta_1}^{n- \sqrt{n}}}_{\check{\theta}}(\omega) \right\} \\
\fl &\le&
1 \times \sup_{\check\theta \notin  U_{\theta_1,\epsilon}}
e^{- n (1- \delta_{n- \sqrt{n}} )D( {\check{\theta}}\|{\theta_1}) }
\end{eqnarray*}
hold. Since $(1- \delta_{n-\sqrt{n}}) \to 1 $, we have
\begin{eqnarray*}
\lim_{n \to \infty}
- \frac{1}{n}\log {\rm P}_{\theta_1}^{M^{w,n}_{ \theta_1}}
\{ T_n \notin U_{\theta_1,\epsilon}\}
=\inf_{\check\theta \notin  U_{\theta_1,\epsilon}}
D( {\check{\theta}}\|{\theta_1})  .
\end{eqnarray*}
Thus, equation (\ref{G26}) is proven.
Then, it implies (\ref{G27}).

Next, we show the weak consistency of $\vec{M^w_{\theta_1}}$.
Assume that 
the true state $\rho_{\theta}$ is not $\rho_{\theta_1}$.
Then, we have 
\begin{eqnarray}
\fl && {\rm P}_{\theta}^{M^{w,n}_{\theta_1}}\{ T_n \notin
U_{\theta,\epsilon_n}\} \nonumber\\
\fl &\le&
{\rm P}_{\theta}^{M_f \times \sqrt{n}} \{ \check{\theta} \notin
U_{\theta,\epsilon_n}\} \nonumber\\
\fl && +
{\rm P}_{\theta}^{M_f \times \sqrt{n}} \{ \check{\theta} \in
U_{\theta,\epsilon_n}\}
\sup_{\check{\theta} \in U_{\theta,\epsilon_n}}
{\rm P}_{\theta}^{E_{\theta_1}^{n- \sqrt{n}}} 
\left \{
e^{n (1- \delta_{n- \sqrt{n}})D( {\check{\theta}}\| {\theta_1})}
{\rm P}^{E_{\theta_1}^{n- \sqrt{n}}}_{\theta_1}
(\omega) \ge
{\rm P}^{E_{\theta_1}^{n- \sqrt{n}}}_{\check{\theta}}(\omega)
\right\},\Label{D3}
\end{eqnarray}
where $\epsilon_n:=
\frac{D(\theta\| \theta_1)}{
2 \left| \Tr \frac{\,d \rho_{\theta}}{\,d \theta}
(\log \rho_{\theta}- \log \rho_{\theta_1})\right|} \delta_n$.
Since $\delta_n= \frac{1}{n^{\frac{1}{5}}}$,
the convergence 
${\rm P}_{\theta}^{M_f \times \sqrt{n}} \{ \check{\theta} \notin
U_{\theta,\epsilon_n}\} \to 0 $ holds.
Also, the relation $U_{\theta,\epsilon_n} \subset 
U_{\theta, \epsilon_{n-\sqrt{n}}}$ holds.
If we can prove
\begin{eqnarray}
\sup_{\check{\theta} \in U_{\theta,\epsilon_n}}
{\rm P}_{\theta}^{E_{\theta_1}^{n}} 
\left \{
e^{n (1- \delta_n)D({\check{\theta}}\| {\theta_1})}
{\rm P}^{E_{\theta_1}^{n}}_{\theta_1}
(\omega) \ge
{\rm P}^{E_{\theta_1}^{n}}_{\check{\theta}}(\omega) \right\}
\to 0, \Label{G31.1}
\end{eqnarray}
we obtain
\begin{eqnarray}
  {\rm P}_{\theta}^{M^{w,n}_{\theta_1}}\{ T_n \notin
U_{\theta,\epsilon_n}\} \to 0. \Label{x1}
\end{eqnarray}
This condition (\ref{x1}) is stronger than the weak consistency condition.
Thus, it is sufficient to show (\ref{G31.1}).

\par From Lemma \ref{L8}, the relations 
\begin{eqnarray}
\fl &&{\rm P}_{\theta}^{E_{\theta_1}^{n}} 
\left \{
e^{n (1- \delta_n)D({\check{\theta}}\| {\theta_1})}
{\rm P}^{E_{\theta_1}^{n}}_{\theta_1}
(\omega) \ge
{\rm P}^{E_{\theta_1}^{n}}_{\check{\theta}}(\omega) \right\} \nonumber\\
\fl &=&
{\rm P}_{\theta}^{E_{\theta_1}^{n}} 
\left \{
\frac{1}{n} 
\left( -\log {\rm P}^{E_{\theta_1}^{n}}_{\check{\theta}}(\omega) 
+
\log {\rm P}^{E_{\theta_1}^{n}}_{\theta_1} (\omega) 
\right) + D( \check\theta \| \theta_1 ) 
\ge  \delta_n D( \check\theta \| \theta_1 ) 
\right\} \nonumber\\
\fl &=&
{\rm P}_{\theta}^{E_{\theta_1}^{n}} 
\Biggl \{
\frac{1}{n} 
\left( -\log {\rm P}^{E_{\theta_1}^{n}}_{\check{\theta}}(\omega) 
+
\log {\rm P}^{E_{\theta_1}^{n}}_{\theta_1} (\omega) 
\right) + \Tr \rho_{\theta} 
( \log \rho_{\check\theta} - \log \rho_{\theta_1}) \nonumber \\
\fl && \ge  \delta_n D( \check\theta \| \theta_1 ) 
+ \Tr (\rho_{\theta} - \rho_{\check\theta})
( \log \rho_{\check\theta} - \log \rho_{\theta_1})
\Biggr\} \nonumber\\
\fl &\le&
{\rm P}_{\theta}^{E_{\theta_1}^{n}} 
\left \{
-\frac{1}{n} 
\log {\rm P}^{E_{\theta_1}^{n}}_{\check{\theta}}(\omega) 
+ \Tr \rho_{\theta} \log \rho_{\check\theta} 
\ge  \delta_n D( \check\theta \| \theta_1 ) 
+ \Tr (\rho_{\theta} - \rho_{\check\theta})
( \log \rho_{\check\theta} - \log \rho_{\theta_1})
\right\} \nonumber\\
\fl &&+
{\rm P}_{\theta}^{E_{\theta_1}^{n}} 
\left \{
\frac{1}{n} 
\log {\rm P}^{E_{\theta_1}^{n}}_{\theta_1} (\omega) 
-\Tr \rho_{\theta} \log \rho_{\theta_1}
\ge  \delta_n D( \check\theta \| \theta_1 ) 
+ \Tr (\rho_{\theta} - \rho_{\check\theta})
( \log \rho_{\check\theta} - \log \rho_{\theta_1})
\right\} \nonumber\\
\fl &\le&
\exp - \Biggl( n \sup_{0 \le t \le 1}
\left(\delta_n D( \check\theta \| \theta_1 ) 
+ \Tr (\rho_{\theta} - \rho_{\check\theta})
( \log \rho_{\check\theta} - \log \rho_{\theta_1})
- \Tr \rho_{\theta}\log \rho_{\check\theta}\right)t \nonumber \\
\fl && \quad - t \frac{(k+1)\log (n+1)}{n}
- \log \Tr \rho_{\theta}\rho_{\check\theta}^{-t}\Biggr) \nonumber \\
\fl && +
\exp - \left( n \sup_{0 \le t} \left(\delta_n D( \check\theta \| \theta_1 ) 
+ \Tr (\rho_{\theta} - \rho_{\check\theta})
( \log \rho_{\check\theta} - \log \rho_{\theta_1})
+ \Tr \rho_{\theta}\log \rho_{\theta_1}\right)t
- \log \Tr \rho_{\theta}\rho_{\theta_1}^{t}\right) 
\nonumber \\
\fl &&\Label{G21}
\end{eqnarray}
hold.
In the following, we assume that $| \theta- \check\theta| \le \epsilon_n$.
Since 
$\epsilon_n=
\frac{D(\theta\| \theta_1)}{
2 \left| \Tr \frac{\,d \rho_\theta}{\,d \theta}
(\log \rho_{\theta}- \log \rho_{\theta_1})\right|} \delta_n$,
we can derive 
$\delta_n D( \check\theta \| \theta_1 ) 
+ \Tr (\rho_{\theta} - \rho_{\check\theta})
( \log \rho_{\check\theta} - \log \rho_{\theta_1})
\le \frac{1}{2}D(\theta\| \theta_1) \delta_n + O(\delta_n^2)$.
Substituting $t=  s \delta_n$, we have
\begin{eqnarray}
\fl && \sup_{\check\theta \in U_{\theta,\epsilon_n}} \frac{1}{n \delta_n^2}
\Biggl( n \sup_{0 \le t \le 1}
(\delta_n D( \check\theta \| \theta_1 ) 
+ \Tr (\rho_{\theta} - \rho_{\check\theta})
( \log \rho_{\check\theta} - \log \rho_{\theta_1})
- \Tr \rho_{\theta}\log \rho_{\check\theta})t \nonumber \\
\fl && - t \frac{(k+1)\log (n+1)}{n}
- \log \Tr \rho_{\theta}\rho_{\check\theta}^{-t}\Biggr) \nonumber\\
\fl &\ge&
\sup_{\check\theta \in U_{\theta,\epsilon_n}} 
\frac{1}{\delta_n^2}
\Biggl(
(\frac{1}{2}D(\theta\| \theta_1) \delta_n + O(\delta_n^2)
- \Tr \rho_{\theta}\log \rho_{\check\theta})s \delta_n
- s \delta_n \frac{(k+1)\log (n+1)}{n} \nonumber \\
\fl && + \Tr \rho_{\theta}\log \rho_{\check\theta} s \delta_n
- \frac{1}{2} ( \Tr \rho_{\theta}(\log \rho_{\check\theta})^2
- (\Tr \rho_{\theta}\log \rho_{\check\theta})^2)
s^2 \delta_n^2 + O(\delta_n^3) \Biggr) \nonumber\\
\fl &\ge& \sup_{\check\theta \in U_{\theta,\epsilon_n}} 
\frac{1}{\delta_n^2}
\Biggl(
\frac{1}{2}D(\theta\| \theta_1) s \delta_n^2 + O(\delta_n^3)
- s \delta_n \frac{(k+1)\log (n+1)}{n} \nonumber \\
\fl && - \frac{1}{2} ( \Tr \rho_{\theta}(\log \rho_{\check\theta})^2
- (\Tr \rho_{\theta}\log \rho_{\check\theta})^2)
s^2 \delta_n^2 + O(\delta_n^3)\Biggr) \nonumber\\
\fl & \to &
\frac{1}{2}D(\theta\| \theta_1) s 
- \frac{1}{2} \left( \Tr \rho_{\theta}(\log \rho_{\theta})^2
- (\Tr \rho_{\theta}\log \rho_{\theta})^2\right)
s^2 \quad (\hbox{ as } n \to \infty) \nonumber\\
\fl &=&
- \frac{1}{2} \left( \Tr \rho_{\theta}(\log \rho_{\theta})^2
- (\Tr \rho_{\theta}\log \rho_{\theta})^2\right)
\left(s - \frac{D(\theta\| \theta_1)}{2( \Tr \rho_{\theta}
(\log \rho_{\theta})^2
- (\Tr \rho_{\theta}\log \rho_{\theta})^2)}\right)^2 \nonumber\\
\fl && 
+ \frac{D(\theta\| \theta_1)^2}{8 ( \Tr \rho_{\theta}
(\log \rho_{\theta})^2
- (\Tr \rho_{\theta}\log \rho_{\theta})^2)}. \nonumber
\end{eqnarray}
Thus, we have
\begin{eqnarray}
\fl && \lim_{n \to \infty}
\sup_{\check\theta \in U_{\theta,\epsilon_n}} 
\frac{1}{n \delta_n^2}
\Biggl( n \sup_{0 \le t \le 1}
(\delta_n D( \check\theta \| \theta_1 ) 
+ \Tr (\rho_{\theta} - \rho_{\check\theta})
( \log \rho_{\check\theta} - \log \rho_{\theta_1})
- \Tr \rho_{\theta}\log \rho_{\check\theta})t \nonumber \\
\fl && - t \frac{(k+1)\log (n+1)}{n}
- \log \Tr \rho_{\theta}\rho_{\check\theta}^{-t}\Biggr) \nonumber\\
\fl &\ge& \frac{D(\theta\| \theta_1)^2}{8 ( \Tr \rho_{\theta}(\log \rho_{\theta})^2
- (\Tr \rho_{\theta}\log \rho_{\theta})^2)} \,> 0 .\Label{G22}
\end{eqnarray}
Also, we obtain
\begin{eqnarray}
\fl && \sup_{\check\theta \in U_{\theta,\epsilon_n}} \frac{1}{n \delta_n^2}
\left( n \sup_{0 \le t}
(\delta_n D( \check\theta \| \theta_1 ) 
+ \Tr (\rho_{\theta} - \rho_{\check\theta})
( \log \rho_{\check\theta} - \log \rho_{\theta_1})
+ \Tr \rho_{\theta}\log \rho_{\theta_1})t
- \log \Tr \rho_{\theta}\rho_{\theta_1}^{t}\right) \nonumber\\
\fl&&\ge \sup_{\check\theta \in U_{\theta,\epsilon_n}} 
\frac{1}{\delta_n^2}
\Biggl(
(\frac{1}{2}D(\theta\| \theta_1) \delta_n + O(\delta_n^2)
+ \Tr \rho_{\theta}\log \rho_{\theta_1})s \delta_n
- \Tr \rho_{\theta}\log \rho_{\theta_1} s \delta_n \nonumber\\
\fl && \quad - \frac{1}{2} ( \Tr \rho_{\theta}(\log \rho_{\theta_1})^2
- (\Tr \rho_{\theta}\log \rho_{\theta_1})^2)
s^2 \delta_n^2 + O(\delta_n^3)\Biggr) \nonumber\\
\fl &&= \sup_{\check\theta \in U_{\theta,\epsilon_n}} 
\frac{1}{\delta_n^2}
\left(
\left(\frac{1}{2}D(\theta\| \theta_1)s  
- \frac{1}{2} \left( \Tr \rho_{\theta}(\log \rho_{\theta_1})^2
- (\Tr \rho_{\theta}\log \rho_{\theta_1})^2\right)
s^2 \right) \delta_n^2 + O(\delta_n^3)\right) \nonumber\\
\fl &&\to
\frac{1}{2}D(\theta\| \theta_1)s  
- \frac{1}{2} ( \Tr \rho_{\theta}(\log \rho_{\theta_1})^2
- (\Tr \rho_{\theta}\log \rho_{\theta_1})^2)
s^2 \quad (\hbox{ as } n \to \infty) .\nonumber
\end{eqnarray}
Therefore,
\begin{eqnarray}
\fl && \lim_{n \to \infty}
\sup_{\check\theta \in U_{\theta,\epsilon_n}} 
\frac{1}{n \delta_n^2}
\left( n \sup_{0 \le t}
(\delta_n D( \check\theta \| \theta_1 ) 
+ \Tr (\rho_{\theta} - \rho_{\check\theta})
( \log \rho_{\check\theta} - \log \rho_{\theta_1})
 + \Tr \rho_{\theta}\log \rho_{\theta_1})t
- \log \Tr \rho_{\theta}\rho_{\theta_1}^{t}\right) \nonumber\\
\fl &&\ge
\frac{D(\theta\| \theta_1)^2}{8 ( \Tr \rho_{\theta}(\log \rho_{\theta_1})^2
- (\Tr \rho_{\theta}\log \rho_{\theta_1})^2)} \,> 0 . \Label{G23}
\end{eqnarray}
Since $n \delta_n^2 \to \infty$,
relation (\ref{G31.1}) follows 
from (\ref{G21}),(\ref{G22}) and (\ref{G23}).

\section{Pinching map and group theoretical viewpoint}\Label{s6}
\subsection{Pinching map in non-asymptotic setting}
In the following, we
prove Lemma \ref{L8} and construct the PVM $E_\theta^n$
after some discussions concerning
the pinching map in the non-asymptotic setting
and group representation theory.
In this subsection, we present some definitions and
discussions of the non-asymptotic setting.

A state $\rho$ is called {\it commutative} with 
a PVM $E(=\{ E_i \})$ on ${\cal H}$ 
if $\rho E_i = E_i \rho $ for any index $i$.
For PVMs $E (=\{ E_i \}_{i \in I}) ,F(=\{ F_j \}_{j \in J})$, 
the notation $E \le  F$ means that
for any index $i \in I$ there exists
a subset $(F/E)_i$ of the index set $J$
such that $E_i = \sum_{j \in (F/E)_i} F_j$.
For a state $\rho$, 
we denote by $E(\rho)$ the spectral measure of $\rho$
which can be regarded as a PVM.
The pinching map ${\cal E}_E$ 
with respect to a PVM $E$ is defined as
\begin{eqnarray}
 {\cal E}_E : \rho \mapsto \sum_{i} E_i \rho E_i , \Label{11.4.1}
\end{eqnarray}
which is 
an affine map from the set of states to itself.
Note that the state ${\cal E}_E(\rho)$ is commutative with a PVM $E$.
If a PVM $F=\{ F_j \}_{j \in J}$ 
is commutative with a PVM $E=\{ E_i \}_{i \in I}$,
we can define the PVM $F \times E= \{ F_j E_i \}_{(i,j)\in I \times J}$,
which satisfies $F \times E \ge E$ and $F \times E \ge F$.
For any PVM $E$,
the supremum of the dimension of $E_i$
is denoted by $w(E)$.
\begin{lem}\Label{thm1}
Let $E$ be a PVM
such that $w(E)\,< \infty$.
If states $\sigma$ and $\rho$ are commutative with the PVM $E$,
and if a PVM $F$ satisfies $E \le F ,E({\sigma}) \le F$,
then we have 
\begin{eqnarray*}
D(\rho \| \sigma ) - \log w(E) \le
D( {\cal E}_F(\rho) \|{\cal E}_F( \sigma) ) \le D(\rho \| \sigma ).
\end{eqnarray*}
\end{lem}
This lemma follows from Lemma \ref{lem1} and Lemma \ref{lem2} below.
\begin{lem}\Label{lem1}
Let $\rho$ and $\sigma$ be states.
If a PVM $F$ satisfies $E(\sigma) \le F$, then
\begin{eqnarray}
D(\rho \| \sigma ) 
= D( {\cal E}_F(\rho) \|{\cal E}_F( \sigma) )+
D ( \rho \| {\cal E}_{F}(\rho ) ) . \Label{pita}
\end{eqnarray}
\end{lem}
\begin{proof}
Since $E(\sigma) \le F$ and 
$F$ is commutative with $\sigma$,
we have $\Tr {\cal E}_{F}(\rho)\log {\cal E}_F(\sigma) = \Tr \rho \log \sigma$.
Since $\rho$ is commutative with $\log \rho$,
we have $\Tr {\cal E}_{F}(\rho)\log \rho = \Tr \rho \log \rho$.
Therefore, we obtain the following:
\begin{eqnarray*}
\fl D( {\cal E}_F(\rho) \|{\cal E}_F( \sigma) )
- D(\rho \| \sigma ) 
&=& \Tr {\cal E}_{F}(\rho) (\log {\cal E}_{F}(\rho) 
- \log {\cal E}_F(\sigma) ) - \Tr \rho (
\log \rho - \log \sigma) \\
\fl &=& \Tr {\cal E}_{F}(\rho) (\log {\cal E}_{F}(\rho) -\log \rho ).
\end{eqnarray*}
This proves (\ref{pita}).
\end{proof} 
\begin{lem}\Label{lem2}
Let $E$ and $F$ be PVMs such that $E \le F$.
If a state $\rho$ is commutative with $E$, we have
\begin{eqnarray}
D ( \rho \| {\cal E}_{F}(\rho) )
\le \log w(E) . \Label{5}
\end{eqnarray} 
\end{lem}
\begin{proof}
Let 
$a_i :=\Tr E_i \rho E_i$ and $\rho_i: = \frac{1}{a_i} E_i \rho E_i$.
Then, we have
$\rho = \sum_{i}a_i \rho_i$,
${\cal E}_{F}(\rho)=
\sum_i a_i {\cal E}_{F}(\rho_i)$,
$\sum_i a_i =1$.
Therefore,
\begin{eqnarray*}
\fl && D ( \rho \| {\cal E}_{F}(\rho) )
= \sum_i \Tr E_i \rho( \log \rho - \log {\cal E}_F(\rho)) 
= \sum_i \Tr E_i \rho E_i 
( E_i\log \rho E_i- E_i\log {\cal E}_F(\rho)E_i) \\
\fl &=& \sum_{i} a_i D ( \rho_i \| {\cal E}_{F}(\rho_i) )
\le \sup_i D ( \rho_i \| {\cal E}_{F}(\rho_i) ) 
= \sup_i \left(
\Tr \rho_i \log \rho_i -
\Tr {\cal E}_{F}(\rho_i) \log {\cal E}_{F}(\rho_i)\right) \\
\fl &\le& - \sup_i \Tr {\cal E}_{F}(\rho_i) \log {\cal E}_{F}(\rho_i)
\le  \sup_i \log \dim E_i
= \log w(E).
\end{eqnarray*}
Thus, we obtain inequality (\ref{5}).
\end{proof}

Let us consider another type of inequality.
\begin{lem}\Label{L9}
Let $E$ be a PVM such that $w(E)\,< \infty$.
If the state $\rho$ is commutative with $E$, and if a PVM $M$ satisfies
that $M \ge E$, we have
\begin{eqnarray}
\rho &\le& {\cal E}_M(\rho) w(E) \Label{G1} \\
\rho^{-t} &\ge& {\cal E}_M(\rho)^{-t} w(E)^{-t}  \Label{G2}
\end{eqnarray}
for $1 \le t \le 0$.
\end{lem}
\begin{proof}
It is sufficient for (\ref{G1})
to show 
\begin{eqnarray}
\rho \le k {\cal E}_M(\rho) , \Label{G3}
\end{eqnarray}
for any state $\rho$ and any PVM $M$ on a $k$-dimensional Hilbert
space ${\cal H}$.
Now, it is sufficient to prove (\ref{G3}) in the pure state case.
For any $\phi, \psi \in {\cal H}$, we have
\begin{eqnarray*}
\lo \langle \psi | k {\cal E}_M (| \phi \rangle \langle \phi| )-
| \phi \rangle \langle \phi | | \psi \rangle 
= k \sum_{i=1}^k \langle \psi | M_i | \phi \rangle \langle \phi |M_i 
|\psi \rangle - \left| \sum_{i=1}^k \langle \psi | M_i | \phi \rangle
\right|^2 \ge 0  .
\end{eqnarray*}
The last inequality follows from Schwartz inequality for
vectors $\{ \langle \psi | M_i | \phi \rangle \}_{i=1}^k$ and 
$\{ 1 \}_{i=1}^k$.
It is well known that the function $u \mapsto - u^{-t}\quad (0 \le t \le
1)$ is an operator monotone function \cite{Bh}.
Thus, (\ref{G1}) implies (\ref{G2}).
\end{proof}
\begin{lem}\Label{L10}
If a PVM $M$ is commutative with a state $\sigma$ and 
$w(M)=1$, we have
\begin{eqnarray}
{\rm P}^M_{\rho} \left\{ \log {\rm P}^M_\sigma(\omega)
\ge a \right\}
\le \exp\left(
- \sup_{0 \le t }
 \left( a t - \log \Tr \rho \sigma^{t} \right) \right) \Label{F2}
\end{eqnarray}
for any state $\rho$.
\end{lem}
\begin{proof} From Markov's inequality, we have
\begin{eqnarray}
 p \left\{ X \ge a \right\} &\le&
\exp - \Lambda_t ( X, p, a)  \Label{L13} \\
 \Lambda_t ( X, p, a) &:=&
at - \log \int e^{t X( \omega) }p(\,d \omega).\nonumber 
\end{eqnarray}
Since $w(M)=1$, the relation $\sum_\omega {\rm P}_{\rho}^M(\omega) 
{\rm P}^M_{\sigma}(\omega)^t=
\Tr {\cal E}_M(\rho){\cal E}_M(\sigma)^t$ holds.
It yields
\begin{eqnarray*}
\Lambda_t ( \log {\rm P}_{\sigma}^{M}, {\rm P}_{\rho}^{M}, a ) 
=
at  - \log \Tr {\cal E}_{M}( \rho)
{\cal E}_{M}( {\sigma})^{t} 
=
at  - \log \Tr \rho {\sigma}^{t}.
\end{eqnarray*}
Thus, we obtain (\ref{F2}).
\end{proof}
\begin{lem}\Label{L11}
Assume that 
$E$ and $M$ are PVMs such that $w(E)\,< \infty, w(M) = 1$ and $M \ge
E$.
If the states $\rho$ and $\rho'$ are commutative with $E$, we have 
\begin{eqnarray}
\fl {\rm P}^M_{\rho} \left\{ - \log {\rm P}^M_{\rho'}(\omega)
\ge a \right\}
\le \exp\left(
- \sup_{0 \le t \le 1}
\left( ( a - \log w(E))t - \log \Tr \rho {\rho'}^{-t} \right)
\right). \Label{L11.1}
\end{eqnarray}
\end{lem}
\begin{proof}
If $ 0 \le t \le 1$, we have
\begin{eqnarray}
\fl  \Lambda_t ( - \log {\rm P}_{\rho'}^{M}, 
{\rm P}_{\rho}^{M}, a ) 
&=&
at  - \log \Tr {\cal E}_{M}( \rho)
{\cal E}_{M}({\rho'})^{-t} 
=
at  - \log \Tr \rho
{\cal E}_{M}({\rho'})^{-t} \nonumber\\
\fl &\ge &
at  -
\log w (E) ^t \Tr \rho
{\rho'}^{-t} \Label{F5}\\
\fl &\ge &
( a  - \log w(E))t - 
\log \Tr \rho {\rho'}^{-t} , \Label{K1}
\end{eqnarray}
where (\ref{F5}) follows from Lemma \ref{L9}.
Therefore, from (\ref{L13}) and (\ref{K1}), we obtain (\ref{L11.1}).
\end{proof}

\subsection{Group representation and its irreducible decomposition}\Label{s31}
In this subsection,
we consider the relation between 
irreducible representations and PVMs for the purpose of
constructing the PVM $E^n_\theta$ and a proof of Lemma \ref{L8}.
Let $V$ be a finite-dimensional vector space 
over the complex numbers $\complex$.
A map $\pi$ from a group $G$ to the generalized linear group
of a vector space $V$ is called a {\it representation} on $V$
if the map $\pi$ is homomorphic, i.e.,
$\pi( g_1 ) \pi (g_2) = \pi ( g_1 g_2 ), ~ \forall g_1 , g_2 \in G$.
The subspace $W$ of $V$ is called {\it invariant} with respect to 
a representation $\pi$ if the vector $\pi (g) w$ belongs to the
subspace $W$ for any vector $w \in W$ and any element $g \in G$.
The representation $\pi$ is called {\it irreducible} if
there is no proper nonzero invariant subspace of $V$
with respect to $\pi$.
Let $\pi_1$ and $\pi_2$ be representations of a group $G$ on 
$V_1$ and $V_2$, respectively.
The {\it tensored} representation $\pi_1 \otimes \pi_2$ of $G$ on $V_1 \otimes V_2$
is defined as
$(\pi_1 \otimes \pi_2) (g) 
= \pi_1  (g) \otimes \pi_2 (g) $,
and the {\it direct sum}
 representation $\pi_1 \oplus \pi_2$ of $G$ on $V_1 \oplus
 V_2$ is also defined as
$(\pi_1 \oplus \pi_2) (g) 
= \pi_1  (g) \oplus \pi_2 (g) $.

In the following,
we treat a representation $\pi$ of a group $G$
on a finite-dimensional Hilbert space ${\cal H}$.
The following fact is crucial in later arguments.
There exists an irreducible decomposition ${\cal H}=
{\cal H}_1 \oplus \cdots \oplus {\cal H}_l$
such that the irreducible components
are orthogonal to one another
if for any element $g \in G$ 
there exists an element $g^* \in G$ such that
$\pi(g)^*= \pi (g^*)$, where $\pi(g)^*$ denotes the adjoint of 
the linear map $\pi(g)$.
We can regard the irreducible decomposition ${\cal H}=
{\cal H}_1 \oplus \cdots \oplus {\cal H}_l$
as the PVM
$\{ P_{{\cal H}_i} \}_{i=1}^{l}$, where $P_{{\cal H}_i}$ denotes
the projection to ${\cal H}_i$.
If two representations $\pi_1$ and $\pi_2$ satisfy the preceding condition,
the tensored representation $\pi_1 \otimes \pi_2$ also 
satisfies it.
Note that in general,
an irreducible decomposition of a representation satisfying the
preceding condition is not unique.
In other words, we cannot uniquely define the PVM from such a representation.

\subsection{Construction of PVM $E_\theta^n$ and
the tensored representation}\Label{s32}
In this subsection, we construct the PVM $E_\theta^n$
after the discussion of the tensored representation.
Let the dimension of the Hilbert space ${\cal H}$ be $k$.
Concerning the natural representation $\pi_{\SL({\cal H})}$ of 
the special linear group $\SL({\cal H})$ on ${\cal H}$,
we consider its $n$-th tensored representation
$\pi_{\SL({\cal H})}^{\otimes n}
:= \underbrace{\pi_{\SL({\cal H})} \otimes \cdots 
\otimes \pi_{\SL({\cal H})}}_n$ on the tensor product space 
${\cal H}^{\otimes n}$.
For any element $g \in \SL({\cal H})$,
the relation $\pi_{\SL({\cal H})}(g)^*=
\pi_{\SL({\cal H})}(g^*)$ holds where the element $g^* \in \SL({\cal
  H})$ denotes the adjoint matrix of the matrix $g$.
Consequently, there exists an irreducible decomposition 
of $\pi_{\SL({\cal H})}^{\otimes n}$ regarded as a PVM
and we denote the set of such PVMs by $Ir^{\otimes n}$.
      
\par From Weyl's dimension formula 
((7.1.8) or (7.1.17) in Weyl \cite{Weyl}and 
Goodman and Wallach \cite{GW}),
the $n$-th symmetric tensor product space is
the maximum-dimensional space in 
the irreducible subspaces with respect to the $n$-th tensored representation 
$\pi_{\SL({\cal H})}^{\otimes n}$.
Its dimension equals the repeated combination $~_{k}H_n$
evaluated by 
$~_kH_{n} =  {n+k-1 \choose k-1} = {n+k-1 \choose n}  
=~_{n+1}H_{k-1}\le (n+1)^{k-1} $.
Thus, any element $ E^n\in Ir^{\otimes n}$ satisfies:
\begin{eqnarray}
w( E^n ) \le (n+1)^{k-1} \Label{s4.2}.
\end{eqnarray}
\begin{lem}\Label{thm2}
A PVM $E^n \in Ir^{\otimes n}$ 
is commutative with the $n$-th tensor product state $\rho^{\otimes n}$
of any state $\rho$ on ${\cal H}$.
\end{lem}
\begin{proof}
If $\det \rho \neq 0$,
this lemma is trivial based on the fact that $\det(\rho)^{-1}
\rho \in \SL({\cal H})$.
If $\det \rho = 0$,
there exists a sequence $\{ \rho_i \}_{i=1}^{\infty}$ 
such that $\det \rho_i \neq 0$ and 
$\rho_i \to \rho$ as $i \to \infty$.
We have 
$\rho_i^{\otimes n} \to \rho^{\otimes n}$ as $i \to \infty$.
Because a PVM $E^n \in Ir^{\otimes n}$ 
is commutative with $\rho_i^{\otimes n}$,
it is also commutative with $\rho^{\otimes n}$.
\end{proof}
\begin{defi}\Label{De2}
We can define the PVM $E^n \times E(\rho^{\otimes n})$ for 
any PVM $E^n \in Ir^{\otimes n}$.
Now we define the PVM $E^n_{\theta}$ satisfying
$w(E^n_\theta)=1$, $E^n_{\theta} \ge E^n \times E(\rho_{\theta}^{\otimes n})$
for a PVM $E^n \in Ir^{\otimes n}$.
Note that the $E^n_{\theta}$ is not unique.
\end{defi}
\noindent{\it Proof of Lemma \ref{L8}:}\quad From 
Lemmas \ref{L10}
and \ref{L11}, (\ref{s4.2}) and the definition of $E^n_{\theta}$,
we obtain Lemma \ref{L8}.
\endproof
\noindent{\it Proof of Lemma \ref{L121}:}\quad From Lemma \ref{thm1}, (\ref{s4.2}) and the definition of $E^n_{\theta}$,
we obtain Lemma \ref{L121}.
\endproof

\section{Large deviation theory for an exponential family}
\Label{s10}
In this section,
we review the large deviation theory for an exponential family.
A $d$-dimensional probability family 
is called an exponential family
if there exist linearly independent real-valued random variables
$F_1, \ldots, F_d$ and a probability distribution
$p$ on the probability space $\Omega$ such that
the family consists of the probability distribution 
\begin{eqnarray*}
p_\theta (\,d \omega):= 
\exp\left( \sum_{i=1}^d \theta^i F_i(\omega) - \psi(\theta)\right)
p(\,d \omega) \\
\psi(\theta):= 
\log \int_\Omega \exp\left( \sum_{i=1}^d \theta^i 
F_i(\omega)\right)p(\,d \omega).
\end{eqnarray*}
In this family, 
the parametric space is given by $\Theta:=
\{ \theta \in \real^d | 0,< \psi(\theta)\,< \infty \}$,
the parameter $\theta$ is called the natural parameter
and the function $\psi(\theta)$ is called 
the potential.
We define the dual potential $\phi(\theta)$ and 
the dual parameter $\eta(\theta)$, called the 
expectation parameter, as
\begin{eqnarray*}
\eta_i(\theta):= \frac{\partial \psi(\theta)}{\partial \theta^i}
= \log \int_\Omega F_i(\omega) p_\theta (\,d \omega)
\Label{14-1}\\
\phi(\theta):= \max_{\theta'} 
\left(\sum_{i=1}^d \theta'^i \eta_i(\theta)
- \psi(\theta')\right).
\end{eqnarray*} From (\ref{14-1}),
we have 
\begin{eqnarray*}
\phi(\theta)= \sum_{i=1}^d \theta^i \eta_i(\theta)
- \psi(\theta).
\end{eqnarray*}
In this family, the sufficient statistics
are given by $F_1(\omega), \ldots, F_d(\omega)$.
The MLE $\hat{\theta}(\omega)$ is given by 
$\eta_i(\hat{\theta}(\omega))= F_i(\omega)$.
The KL divergence $D(\theta\| \theta_0):=D(p_{\theta}\| p_{\theta_0})$ 
is calculated by
\begin{eqnarray*}
\fl D(\theta\| \theta_0)
&=&  \int_{\Omega} \log \frac{p_\theta(\omega)}
{p_{\theta_0}(\omega)}p_\theta(\,d \omega)
=  \int_{\Omega}
\sum_i (\theta^i- \theta^i_0 )F_i(\omega)+ \psi(\theta_0)-\psi(\theta)
p_\theta(\,d \omega)\\
\fl &=&  \sum_i (\theta^i- \theta^i_0 )\eta_i(\omega)+ 
\psi(\theta_0)-\psi(\theta)
= \phi(\theta) +\psi(\theta_0) - \sum_i \theta^i_0 \eta_i(\omega) \\
\fl &=& \max_{\theta'}
\left( \sum_i \theta'^i\eta_i(\theta)- 
\psi(\theta') \right)+ \psi(\theta_0)
- \sum_i \theta_0^i \eta_i(\theta) \\
\fl &=& \max_{\theta'}
\sum_i (\theta'^i- \theta'^i_0)\eta_i(\theta)- 
\log \int_\Omega \exp \left(\sum_i (\theta^i- \theta^i_0 )
F_i(\omega)\right)p_\theta(\,d \omega).
\end{eqnarray*}

Next, we discuss the $n$-i.i.d.\ extension of
the family $\{ p_\theta | \theta \in \Theta\}$.
For the data $\vec{\omega}_n:= (\omega_1 , \ldots, \omega_n)
\in \Omega^n$,
the probability distribution
$p^n_\theta(\vec{\omega}_n):= 
p_\theta(\omega_1) \ldots p_\theta(\omega_n)$
is given by
\begin{eqnarray*}
p^n_\theta(\vec{\omega}_n)
= \exp \left(n \sum_i \theta^i F_{n,i}(\vec{\omega}_n)- n \psi(\theta)\right)
p^n (\,d \vec{\omega}_n)\\
p^n (\,d \vec{\omega}_n):=
p(\,d \omega_1) \ldots p(\,d \omega_n) \\
F_{n,i}(\vec{\omega}_n):= 
\frac{1}{n}\sum_{k=1}^n F_i(\omega_k).
\end{eqnarray*}
Since the expectation parameter of the probability family
$\{ p^n_\theta| \theta \in \Theta\}$
is given by $n \eta_i(\theta)$,
the MLE $\hat{\theta}_n(\vec{\omega}_n)$
is given by
\begin{eqnarray}
n \eta_i(\hat{\theta}_n(\vec{\omega}_n))
= n F_{n,i}(\vec{\omega}_n). \Label{14-2}
\end{eqnarray}
Applying Cram\'{e}r's Theorem \cite{Buck}
to the random variables $F_1, \ldots, F_d$
and the distribution $p_{\theta_0}$,
for any subset $S \subset \real^d$
we have 
\begin{eqnarray*}
\fl  \inf_{\eta \in \overline{S}}
\sup_{\theta' \in \real^d}
\left( \sum_i \theta'^i(\eta_i - {\rm E}_{\theta_0}(F_i))
- \psi_{\theta_0}(\theta') \right)
&\le&
\lim_{n \to \infty} \frac{-1}{n}
\log p^n_{\theta_0} \{ \vec{F_n} \in S\} \\
\fl &\le&
\inf_{\eta \in {\rm int} S}
\sup_{\theta' \in \real^d}
\left( \sum_i
\theta'^i(\eta_i - {\rm E}_{\theta_0}(F_i))
- \psi_{\theta_0}(\theta') \right),
\end{eqnarray*}
where
\begin{eqnarray*}
{\rm E}_{\theta_0}(F_i)):= 
\int_\Omega F_i(\omega) p_\theta(\,d \omega) \\
\psi_{\theta_0}(\theta):=
\int_\Omega
\exp \left( \sum_i \theta^i F_i (\omega)\right) p_\theta(\,d \omega) \\
\vec{F_n}(\vec{\omega}_n):= 
( F_{n,1}(\vec{\omega}_n), \ldots, F_{n,d}(\vec{\omega}_n)),
\end{eqnarray*}
and ${\rm int} S$ denotes the interior of $S$,
which is consistent with $(\overline{S^c})^c$.
Since 
\begin{eqnarray*}
\fl \sup_{\theta' \in \real^d} \left(
\sum_i
\theta'^i(\eta_i - {\rm E}_{\theta_0}(F_i)) 
- \psi_{\theta_0}(\theta')\right)
= \sup_{\theta' \in \real^d}\left(
\sum_i \theta'^i(\eta_i - \eta_i(\theta_0))
- \psi(\theta') \right)+ \psi(\theta_0) =
D(\theta\|\theta_0)
\end{eqnarray*}
and the map $\theta \mapsto D(\theta\|\theta_0)$
is continuous,
it follows from (\ref{14-2}) that
\begin{eqnarray*}
\lim_{n \to \infty} \frac{-1}{n}
\log p^n_{\theta_0} \{ \hat{\theta}_n \in\Theta' \}
=
\inf_{\theta \in\Theta'}
D(\theta\|\theta_0)
\end{eqnarray*}
for any subset $\Theta' \subset \Theta$,
which is equivalent to (\ref{Dec5}).
Conversely,
if an estimator $\{ T_n (\vec{\omega}_n)\}$
satisfies the weak consistency
\begin{eqnarray*}
\lim_{n \to \infty}
p_\theta^n \{ \| T_n (\vec{\omega}_n) - \theta \|
\,> \epsilon \} \to 0 , \quad
\forall \epsilon \,> 0 , \forall \theta \in \Theta,
\end{eqnarray*}
then, similarly to (\ref{a21}), we can prove 
\begin{eqnarray*}
\lim_{n \to \infty} \frac{-1}{n}
\log p^n_{\theta_0} \{  T_n (\vec{\omega}_n) \in\Theta' \}
\le \inf_{\theta \in\Theta'}
D(\theta\|\theta_0).
\end{eqnarray*}
Therefore, we can conclude that
the MLE is optimal in the large deviation sense 
for exponential families.

\section{Estimation of spectrum for unitary invariant family}\Label{ap15}
Suppose that a multi-parametric quantum state family ${\cal S}$
satisfies 
\begin{eqnarray*}
U \rho U^* \in {\cal S} , \hbox{ for }
\forall \rho \in {\cal S},
\end{eqnarray*}
and that the vector ${\bf p}(\rho)= (p_1(\rho), \ldots, p_d(\rho))$
satisfies $p_i(\rho) \ge p_{i+1}(\rho) $,
where $d$ is the dimension of ${\cal H}$.
Keyl and Werner's estimator 
$\vec{M}_{KW}= \{ M^n_{KL}\}$ satisfies 
\begin{eqnarray}
\lim\frac{-1}{n}\log 
{\rm P}^{M^n_{KW}}_{\rho^{\otimes n}}\{ \hat{\bf p} \in {\cal R} \}
= 
\inf _{{\bf p} \in {\cal R}} D( {\bf p} \| {\bf p}(\rho)),
\Label{1a}
\end{eqnarray}
where ${\cal R}$ is a subset consisting of $d$-nomial
distributions\cite{KW}.
Conversely, if a sequence of estimators $\vec{M}=\{M^n\}$ satisfies 
\begin{eqnarray*}
{\rm P}^{M^n}_{\rho^{\otimes n}} 
\left\{ \left\|\hat{\bf p}  - {\bf p}(\rho) \right\|  \,> 
\epsilon \right\}
\to 0 , \quad \forall \epsilon \,> 0, \forall \rho \in {\cal S},
\end{eqnarray*}
then we can show 
\begin{eqnarray}
\limsup \frac{-1}{n} \log 
{\rm P}^{M^n}_{\rho^{\otimes n}} \{ \hat{\bf p}  \in {\cal R}\}
\le \inf_{{\bf p}(\sigma) \in {\cal R} }
D(\sigma \|\rho)\Label{1b}
\end{eqnarray}
by a similar way to (\ref{a21}).
Since 
\begin{eqnarray*}
\min_{U:\hbox{unitary}}
D(U\sigma U^* \| \rho) = D({\bf p}(\sigma) \|{\bf p}(\rho)),
\end{eqnarray*}
the RHS of (\ref{1b}) equals the RHS of (\ref{1a}).
Therefore, Keyl and Werner's estimator $\vec{M}_{KW}$
is optimal in the sense of large deviation.
Now, we consider a parametric subspace 
$\{ {\bf p}_\theta | \theta \in \Theta\}$ of $d$-nomial
distributions.
Assume that ${\bf p}(\rho) = {\bf p}_{\theta_0}$,
then 
\begin{eqnarray}
\lim_{\epsilon \to 0} \frac{1}{\epsilon^2}
\inf_{\|\theta - \theta_0 \| \,> \epsilon }
D({\bf p}_{\theta}\|{\bf p}_{\theta_0})
=
\frac{1}{2}\min_{\|\xi\|=1 }J_{\theta;i,j}\xi_i \xi_j,
\Label{1c}
\end{eqnarray}
where $J_{\theta;i,j}$ is Fisher information matrix of 
$\{{\bf p}_\theta | \theta \in \Theta\}$.
Since the convergence of the LHS of (\ref{1c})
is uniform and the RHS of (\ref{1c}) 
is continuous for $\theta$,
the bound of the weak consistency coincides with
the bound of the strong consistency.

\end{document}